\documentclass{article}
\usepackage{emulateapj}

\def\lea{\mathrel{<\kern-1.0em\lower0.9ex\hbox{$\sim$}}}
\def\gea{\mathrel{>\kern-1.0em\lower0.9ex\hbox{$\sim$}}}
\newcommand{\NV}{$\mbox{N~V}~\lambda1240$~}
\newcommand{\SiIV}{$\mbox{Si~IV}~\lambda1400$~}
\newcommand{\CIV}{$\mbox{C~IV}~\lambda1550$~}

\begin{document}

\slugcomment{Accepted for Publication in the Astrophysical Journal}

\title{The Stellar Content of Nearby Star-Forming Galaxies. III. Unravelling the Nature of the Diffuse Ultraviolet Light\altaffilmark{1}}

\author{\sc Rupali~Chandar\altaffilmark{2}, 
  Claus~Leitherer\altaffilmark{2}, 
  Christy~A.~Tremonti\altaffilmark{3},
  Daniela~Calzetti\altaffilmark{2},
  Alessandra~Aloisi\altaffilmark{2,4},
  Gerhardt~R.~Meurer\altaffilmark{5},
  Duilia~de Mello\altaffilmark{6,7}
}

\altaffiltext{1}{Based on observations with the NASA/ESA {\it Hubble
Space Telescope}, obtained at the Space Telescope Science Institute, which
is operated by the Association of Universities for Research in Astronomy,
Inc. under NASA contract NAS5-26555. }
\altaffiltext{2}{Space Telescope Science Institute, 3700 San Martin Drive, Baltimore, Maryland 21218; Electronic Address: rupali@stsci.edu}
\altaffiltext{3}{Steward Observatory, 933 N. Cherry Ave., Tucson, AZ, 85721}
\altaffiltext{4}{On assignment from the Space Telescope Division of ESA}
\altaffiltext{5}{Johns Hopkins University, 3400 N.~Charles Street, Baltimore, MD 21218}
\altaffiltext{6}{Laboratory for Astronomy and Solar Physics, Code 681, Goddard Space Flight Center, Greenbelt, MD 20771}
\altaffiltext{7}{Department of Physics, Catholic University of America, 620 Michigan Avenue, Washington, DC 20064}

\begin{abstract} 

We investigate the nature of the diffuse intra-cluster ultraviolet
light seen in twelve local starburst galaxies, using long-slit
ultraviolet spectroscopy obtained with the Space Telescope Imaging
Spectrograph (STIS) aboard the {\it Hubble Space Telescope\/} (\emph{HST}).
We take this faint intra-cluster light to be the field in each galaxy,
and compare its spectroscopic signature with STARBURST99 evolutionary
synthesis models and with neighboring star clusters.  Our main result
is that the diffuse ultraviolet light in eleven of the twelve
starbursts lacks the strong O-star wind features that are clearly
visible in spectra of luminous clusters in the same galaxies.  The
difference in stellar features dominating cluster and field spectra
indicate that the field light originates primarily from a different
stellar population, and not from scattering of UV photons leaking out
of the massive clusters.  A cut along the spatial direction of the UV
spectra establishes that the field light is not smooth, but rather
shows numerous ``bumps and wiggles.''  Roughly 30--60\% of these
faint peaks seen in field regions of the closest ($<4$~Mpc) starbursts
appear to be resolved, suggesting a contribution from superpositions
of stars and/or faint star clusters.  Complementary WFPC2 \emph{UVI}
imaging for the three nearest target galaxies, NGC~4214, NGC~4449, and
NGC~5253 are used to obtain a broader picture, and establish that all
three galaxies have a dispersed population of unresolved, luminous
blue sources.  Because the field spectra are dominated by B stars, we
suggest that the individual sources observed in the WFPC2 images are
individual B~stars (rather than O~stars), or small star clusters.  We
consider several scenarios to understand the lack of observed massive
stars in the field, and their implications for the origin of the field
stellar population.  If the field stellar populations formed {\it in
situ\/}, the field must either have an IMF which is steeper than
Salpeter ($\alpha \sim-3.0~\mbox{to}-3.5$), or a Salpeter slope with
an upper mass cutoff of 30--$50~M_{\odot}$.  If star formation occurs
primarily in star clusters, the field could be composed of older,
faded clusters, and/or a population which is coeval with the luminous
clusters but lower in mass.  We use these benchmark populations to
place constraints on the field stellar population origin.  Although
the field probably includes stars of different ages, the UV light is
dominated by the youngest stellar populations in the field.  If the
field is composed of older, dissolving clusters, we estimate that star
clusters (regardless of mass) need to dissolve on timescales
7--10~Myr to create the field. If the field is composed of young
clusters which fall below the detection limit of individual sources in
our spectroscopy, they would have to be several hundred solar masses
or less, in order to be deficient in O stars, despite their extreme
youth.  

\end{abstract}
 

\section{INTRODUCTION}

Nearby starburst galaxies are fundamental testing grounds for key
questions related to both star formation and galaxy evolution, as well
as the interplay between galaxies and their surroundings.  A detailed
understanding of the properties of local starbursts is also critical
to further examine star forming galaxies at high redshift, since the
star formation rates of these distant, young galaxies are most similar
to those found in local starbursts (e.g., Steidel et~al.\ 1996).
However, we are still learning about the stellar content and
properties of local UV-bright starbursts.  Meurer et~al.\ (1995) showed
that nearby starbursts are irregular structures consisting of
diffusely distributed light interspersed with prominent star clusters,
including both compact objects (with half-light radii $\lea 10$~pc),
and more extended OB associations.  They estimate that clusters only
contribute $\sim20$\% of the light seen in ultraviolet images of
starburst galaxies (Meurer et~al.\ 1995; Maoz et~al.\ 1996).  Thus the
diffusely distributed light actually dominates the starburst,
comprising roughly 80\% of the total in the UV.
To account for the observed ``bimodal'' nature of
stellar populations in starbursts, Meurer et~al.\ (1995) suggested that
two modes of star formation occur in high intensity bursts---prominent 
cluster formation and dominant diffusely distributed star formation.

Several explanations besides star formation occurring in a ``cluster''
mode and a ``field'' mode can explain the observed dichotomy of
stellar properties in local starbursts.
Currently one popular scenario suggests that the diffuse UV field
light is created via dissipation of aging star clusters, which
contribute their remaining stars to the field.  There are several
works which indirectly support this scenario (e.g.,
Tremonti et~al.\ 2001; Lada \& Lada 2003; Fall, Chandar, \& Whitmore 2004).

The stellar content of nearby starburst galaxies have been studied in
more detail using multiband optical WFPC2 photometry.  In addition to
star clusters, color$-$magnitude diagram (CMD) analysis of the field
star population suggests two distinct overall populations.  Greggio et~al.\ 
(1998) and Harris et~al.\ (2001) find evidence that the resolved,
diffuse stellar populations in NGC~1569 and M83 are older than the
clusters.  Tremonti et~al.\ (2001) compared UV spectroscopy of both
resolved clusters and the diffuse field light in the nearby starburst
NGC~5253, with stellar evolutionary models.  They established that the
diffuse intra-cluster light in NGC~5253 has a spectrum lacking the
strong O-star wind features which are clearly seen in a number of the
cluster spectra in this galaxy.
All of these observations are consistent with a scenario where the
field is composed of dissolved star clusters.  A possible
counter-example, which supports the concept of stars forming early on
in both clusters {\it and\/} in the field, is the Tosi et~al.\ (2001)
study of NGC~1705.  They found that in order to reproduce the CMD of
point sources in this post-starburst galaxy, a 2--3~Myr
{\it field\/} star population was required.

However, other scenarios can also plausibly explain the presence of
diffuse UV field light in starbursts.  For example, continuous star
formation in galaxies could result in a population of B~stars, which
originally formed in clusters over the last few 100~Myr, and dispersed
throughout the galaxy once the clusters dissolved.  A second
possibility is that a fraction of UV photons originating in hot stars
embedded in HII regions leak out and scatter off dust grains. In 
30~Doradus, UIT observations suggest that scattered light is the
mechanism responsible for the UV field (Cheng et~al.\ 1992).  In
He~2-10 however, a much more distant starburst, we were able to
establish that scattered light cannot be the dominant mechanism
responsible for the observed UV field (Chandar et~al. 2003).  Barth 
et~al.\  (1995) suggested a third possibility, that the diffuse light in
nuclear starburst rings may be dominated by (lower mass), unresolved
star clusters.  A fourth possible explanation is that individual
massive stars may be forming in relative isolation in the field, as
suggested for the nuclear regions of M51 (Lamers et~al.\ 2002).
Finally, in our recent work on the stellar content of He~2-10, we
established that both the clusters and intra-cluster regions show
signatures of massive stars.  From our detailed simulations, we
suggest that the field population which formed coevally with the
neighboring compact clusters ($\sim$4--5~Myr ago) may come from less
spatially concentrated ``scaled OB assocations'' (SOBAs), which are
not lacking in the most massive stars, rather than faint, low mass
coeval compact star clusters.

Is He~2-10 a unique case, where we are ``catching'' this galaxy early
during a well coordinated burst?  Do other galaxies have field stellar
populations which appear older than those of neighboring clusters,
such as found in NGC~5253?  Here we extend our previous studies of the
diffuse intra-cluster light in local starbursts using longslit STIS UV
spectroscopy.  Ten galaxies from our \emph{HST} program GO-9036 have high
enough S/N in the field to allow us to study their stellar
composition.  To these, we add the published results of He~2-10 and
NGC~5253 for a more complete look at the nature of diffuse UV light in
starburst galaxies.  

This paper is organized as follows: $\S2$ provides background
information on the galaxy sample and describes the observations, basic
data reduction, and definition of the field regions; $\S3$ describes
reduction and basic analysis of complementary archival WFPC2 images
for three of the closest target galaxies (NGC~4214, NGC~4449, and
NGC~5253); $\S4$ derives the stellar content of the field by comparing
with STARBURST99 models; and $\S5$ takes
advantage of spatial information as well as the spectra, in order to
constrain the origin of the UV field light in our sample.  Finally, in
$\S6$ we summarize the main conclusions of this work.

\section{THE SAMPLE, OBSERVATIONS AND DATA REDUCTION}

\subsection{Galaxy Sample}

Our \emph{HST} program, GO-9036, was designed to obtain a homogeneous far-
and near- UV spectroscopic dataset of 18 nearby starburst galaxies.
The target galaxies span a broad range of morphologies, chemical
compositions, and luminosities.  The primary targets in each galaxy
were one to three luminous clusters (either compact or giant H~II
regions); however, the projected $25\arcsec \times 0.2\arcsec$ STIS
long-slit also covers fainter, serendipitous objects (which could be
stars or clusters), and regions of diffuse UV emission where no
obvious sources are seen in available archival FOC and WFPC2 images.
The galaxy sample, distances, and area projected in the slit are
compiled in Table~\ref{table:sample}.  Additional details for the
galaxy sample and observations (including slit locations) are
presented in the first paper in our series (Chandar et~al.\ 2005a;
hereafter paper~I).  Analysis of the cluster content is the focus of
the second paper (Chandar et~al.\ 2005b; hereafter paper~II).  In this
investigation, we analyse the diffuse UV emission seen between
clusters in ten target galaxies.  These are the galaxies where the
signal-to-noise (S/N) of the extracted field spectrum is sufficiently
high ($\gea10$) to quantify the dominant stellar population.  To these
we add the results of our previous work in He~2-10 (Chandar et~al.\ 
2003) and NGC~5253 (Tremonti et~al.\ 2001), for a comprehensive
study of diffuse UV emission in 12 local starburst galaxies.

\subsection{Data Reduction}

In order to combine multiple aligned exposures of the same target, we
added together the flatfielded two-dimensional images (there were no
shifts between multiple galaxy spectra taken for GO-9036), updated the
exposure time in the headers to reflect the total integration time,
and reprocessed through the CALSTIS pipeline.  This pipeline rebins
the spectra and provides a global detector linearity correction, dark
subtraction, flat fielding, wavelength calibration and conversion to
absolute flux units.  For subsequent analysis on the field regions we
utilized these fully calibrated rectified two dimensional images
reprocessed through the CALSTIS pipeline, which have a linear
wavelength scale and uniform sampling in the spatial direction.
Further processing steps on the extracted one dimensional spectra
included deredshifting (values taken from NED) and rebinning to the
wavelength scale of the STARBURST99 models
(0.75~\AA~$\mbox{pixel}^{-1}$) utilized for comparison with the
observations.  We then corrected the extracted spectra for strong
geocoronal emission at $\mbox{Ly}\alpha$ and OI $\lambda 1302$ by
assuming that an outer portion from the two dimensional spectrum,
which is typically located near the edge of the observable galaxy,
represents a purely geocoronal spectrum near the $\mbox{Ly}\alpha$ and
OI $\lambda 1302$ features.  For each galaxy, the spectrum extracted
from near the end of the slit is typically at least $10\times$ weaker
in continuum flux then the field regions, and displays very weak
spectral features (if they were observable at all), indicating that
these are suitable for correcting geocoronal features.

\subsection{Extraction of the Field}

In order to study the stellar content of the galaxies in our sample,
we first detected objects (which could be stars or clusters) from
profiles taken along the slit (700 columns were collapsed along the
spatial direction).  Our detection procedure is analogous to those
generally used for object detection in images.  Objects were defined
as peaks along the spatial direction which reached a level at least
five times higher than the standard deviation ($\sigma$) plus the
local background level (a second category, faint objects, which have
3--$5\sigma$ detections are also studied in paper~II).  This technique
ensures that we are detecting real sources and not random surface
brightness fluctuations.  In principle, detected objects can be either
stellar clusters or individual stars.  However, individual stars
should be considered as part of the diffuse UV population.  Therefore,
it is very important to assess whether our object detection algorithm
is detecting individual stars, since we do not want to exclude
individual~O and/or B~stars from our definition of the field.  Based
on the depth of our data and detection limits, the possibility of
excluding individual massive stars is only a concern for the closest
galaxies, those within $\sim$5~Mpc.  Because clusters dim as they age,
the total luminosities of clusters and individual O/B supergiants can
overlap, making it impractical to use a simple luminosity cut to
separate stars and clusters.  A further complication is that the
narrow slit used in this study often does not include all of the light
from an object, either because the object is extended, and/or because
it is not centered in the slit.  Thus it is important to consider the
{\it total\/} intrinsic luminosities of detected sources, and not just
the light falling in the slit.  To address this concern, we use a
combination of object luminosities (from paper~II) measured at
1500~\AA, which are strict lower limits, and size information (from
profiles along the STIS slit and sometimes available WFPC2 images) to
assess whether we are indeed detecting individual luminous stars.

Here we use NGC~5253 as an example to examine the nature of detected
sources, since the FUV spectrum for this galaxy is the deepest of the
three closest targets.  An O~supergiant has an intrinsic spectral
luminosity density at 1500~\AA, $L_{1500}$, of
$\sim5\times10^{35}~\mbox{erg~s}^{-1}~\mbox{\AA}^{-1}$.  This is
similar to the light falling in the STIS slit for the weakest detected
objects (see e.g., objects~3 and 11 in Tremonti et~al.\ 2001).
However, the spatial profiles for these objects are clearly resolved
(as are the profiles for basically all detected objects in NGC~5253
and NGC~4214, and most objects in NGC~4449), and the slit location
suggests that much of the flux from object~3 in NGC~5253 has been
missed.  In fact, a comparison of the total $L_{1500}$ from available
F170W WFPC2 images and STIS spectroscopy (see Table~1 in Tremonti 
et~al. 2001) shows that for the four brightest sources, the $L_{1500}$
flux measured from imaging where all of the light from an object is
accessible, is higher by factors of 4--15 than the flux measured from
STIS spectroscopy.  Similar arguments apply for the objects detected
in NGC~4214 and NGC~4449.  We conclude that the detected objects in
galaxies closer than 5~Mpc are not individual luminous stars, but
stellar clusters or groups of stars.  For more distant galaxies object
fluxes are too high to come from individual stars.  Therefore, we will
refer to detected sources in our STIS spectra as clusters.  In our
analysis, we make an arbitrary distinction in each galaxy between
bright clusters ($\mbox{S/N}>10$) and faint clusters (typically $4\leq
\mbox{S/N} \leq 8$).  These are discussed further in $\S5.2.2$.

We defined the field regions for each galaxy in two ways: (1)~regions
within $3\sigma$ of the local background, and (2)~the entire galaxy
minus the 1--3 targeted luminous clusters (those used to determine
the slit orientation).  The spatial pixels defined as belonging to the
field were then summed to give a one-dimensional field spectrum.  Note
that in regions where the background changes rapidly along the slit,
it can be difficult to determine the appropriate local level.  To
address this issue and any lingering concerns that our first
definition eliminates small groups of individual O or B~stars from the
extracted field, we analysed both extracted field spectra for each
galaxy.  We minimized contamination from luminous clusters by avoiding
light from the wings of their profiles.  Spatial profiles, along with
field regions according to our first definition are presented in
Figure~1.

Note that a number of galaxies, particularly those with a large number
of clusters in the slit (e.g., NGC~3310 and NGC~4449), also have many
``bumps and wiggles'' in their field portions (those within $3\sigma$
of the local background).  In fact, in most galaxies, particularly
those which are nearby, star clusters are not embedded in a smooth
background of UV light.  The field always shows some structure.  Is
this structure due to variable reddening or discrete stellar
populations?  In $\S5.4$ we discuss the results of measuring the
spatial extent or size of numerous faint ``peaks'' visible in
starburst field regions through our slits.  We find in general that a
significant fraction of these peaks are unresolved, and those which
are resolved have sizes which are comparable to luminous clusters
observed in each galaxy.  Additionally, we were able to match a few of
the brighter peaks observed in the STIS field regions with discrete,
faint sources in deeper archival WFPC2 images.  Taken together, these
points suggest that at least some of the faint peaks seen along the
spatial direction are actually discrete stellar populations (possibly
a few stars or faint compact clusters), rather than variations in
extinction or random fluctuations in the background level.
\newpage

\subsection{Selection Effects/Biases}

We note that the precise locations defined as the ``field'' in each
galaxy depend on a number of factors, such as total exposure time,
galaxy distance, and local background levels.  Thus the ``field'' as
we have defined it, is subject to a number of selection
effects/biases.  For example, deeper observations might allow for
additional small bumps and wiggles in the spectrum to be added to the
``object'' list, as their S/N is increased.  However, because we have
collapsed a large number of pixels along the spatial direction, our
definitions are somewhat robust against the depth of the observations.
Our definition of ``field'' is also distance dependent in the sense
that we are unable to uniformly define a minimum object mass at a
given age which is selected as a cluster.  More distant galaxies are
more likely to have higher mass clusters hiding in the field.  If some
of these clusters are very young, than we would expect the signature
of massive stars in the extracted field spectra to increase with
galaxy distance.  Potentially, some of the less massive objects that
we have defined as clusters (this only applies to galaxies closer than
$\sim$5~Mpc) may be random superpositions of several field O~stars, or
possibly ``scaled OB associations'' or SOBAs (Hunter 1999).

Our second
extraction of the field, where we have only excised targeted clusters,
allows us to directly probe this issue.  In $\S5.2.2$ we present
results comparing the stellar content in the field and cluster spectra
with instantaneous burst STARBURST99 models.  Although these models
are not a good physical representation of the field, they do provide a
simple way to quantify the massive star content.  We found virtually
no difference in the best fit ages, and hence in the massive star
content, in our two definitions of the field for each galaxy.

\subsection{Spatial Profiles Along the Slit}

Figure~1 shows the integrated 1250--1700~\AA\ flux along the spatial
direction of our longslit spectra.  In addition to allowing us to
define clusters/field portions, these spatial profiles give some
insight into the stellar variations across each galaxy.  In
Table~\ref{table:frac} we present the total fraction of FUV light from
star clusters observed in each pointing.  We summed up light from
clusters (along the spatial direction) and compared with the total
light along the slit.  The fraction of light in the field is assumed
to be that not found in clusters.  We derive values for the fraction
of light in the field portions of local starburst galaxies ranging
from roughly 25\% to 70\%.  However, our slit locations targeted
UV-bright clusters, and whenever possible orientations were chosen to
maximize the number of such clusters in the slit.  Thus the particular
locations observed in each galaxy are biased {\it against\/} the field.
The field fractions presented in Table~\ref{table:frac} represent
lower limits to the field contribution (and an upper limit to the
cluster contribution) in the FUV, and are roughly consistent with the
$\sim$50--80\% field contribution derived by Meurer et~al.\ (1995) when
considering nearby starbursts in their entirety.  Below, we briefly
describe the spatial cuts for each sample galaxy.

{\it Mkn33:} There are field regions on either side of, and between, 
two luminous UV regions which contain multiple clusters.

{\it He~2-10:} We previously made a detailed study of starburst region
A (Chandar et~al.\ 2003). Along the spatial direction of our slit,
there are five individual compact clusters on an elevated background
at UV-optical wavelengths.  The nature of this diffuse field light was
investigated, and found to include a significant population of massive
stars.  This population could arise in a large number of lower mass
compact clusters, or from diffuse SOBAs. Because the density of
putative lower mass clusters would have to be very large, we concluded
that SOBAs formed coevally with the compact clusters were likely the
main stellar contributor to the massive star-rich field.

{\it NGC~1741:} 
The field region in this galaxy includes a broad, low level component
(see Figure~1).

{\it NGC~3125:} There is some diffuse UV emission observed between
three bright clusters.  The field includes a number of faint peaks.

{\it NGC~3310:} The spatial cut of this spiral galaxy shows a
multitude of peaks.  Based on the definition given in $\S2.3$, we
detect ten clusters.  The field region includes a number of very faint
peaks.

{\it NGC~4214:} In Figure~2 we show an enlargement of the spatial
profile along the STIS slit for part of the field in NGC~4214.  Faint,
individual peaks are clearly seen.  The field appears qualitatively
similar to that in NGC~5253, where a number of bumps and wiggles are
seen on a broad, elevated UV background.

{\it NGC~4449:} This galaxy is similar to NGC~3310 in the spatial
direction, where a relatively large number (12) of individual clusters
are found, and the field includes additional, faint peaks.

{\it NGC~4670:} This galaxy has a profile similar to that seen in
He~2-10---there is a somewhat peaked plateau of diffuse emission, with
bright clusters superposed.  There is however, slightly more
structure seen in the field of NGC~4670 than that in He~2-10.

{\it NGC~5253:} This galaxy was studied by Tremonti et~al.\ (2001). 
There is an elevated UV background, with a number of small peaks (included
in the field), and larger peaks (defined as individual clusters).

{\it NGC~5996:} In addition to the targeted central cluster, a couple
of other fainter clusters and field regions are observed in the slit.

{\it NGC~7552:} There are two bright clusters on an elevated background in
the slit, with some field portions in-between.

{\it TOL1924-416:} There is one very luminous cluster, plus one
fainter one.  The field consists of a number of low level bumps
and wiggles.

\section{COMPLEMENTARY ARCHIVAL WFPC2 IMAGES}

\subsection{Data and Reduction}

To further constrain properties of the field regions studied in this
work, we downloaded available archival WFPC2 images for our target
galaxies.  These allow the study of the stellar
populations along the slit at different wavelengths, as well as
providing a broader view of the stellar populations in our target
galaxies, and not just the pointings targeted by our STIS longslit
spectra.  Because much of the available archival data is not very
deep, or does not include observations in several filters, we restrict
the WFPC2 image analysis to the three closest galaxies, NGC~4214,
NGC~4449, and NGC~5253, where individual stars can easily be detected,
and the UV spectra are least subject to issues related to distance
bias.  These three target galaxies all have observations in filters
which can be converted to Johnson-Cousins~$U$, $V$, and~$I$.  In
Table~\ref{table:phot}, we compile information on these galaxies and
on the archival WFPC2 observations used in this work.
Images were processed through the \emph{HST} on-the-fly calibration pipeline
(which automatically selects the most up-to-date calibration files
during processing).  Multiple images were combined to eliminate cosmic
rays.

We ran the object finding algorithm SEXTRACTOR (Bertin \& Arnouts
1996), to locate all objects (individual stars, clusters, and a
handful of background galaxies) in the F336W images (the closest match
to the wavelength range of our STIS observations).  We used a
threshold of $3\sigma$ above the local background level, which upon
visual inspection appeared to detect all obvious sources.  In general,
a detection which is the mean of the local background plus 3 times the
standard deviation ($\sigma$) of this background, eliminates random
surface brightness fluctuations, and leads to detection of actual
objects.  The F336W filter has a significant red leak, causing
a fraction of an object's flux longward of 4000~\AA\ to be detected in
this filter. Biretta et~al.\ (2000) show that the red leak in this filter is
generally $\lea5$\% of the total flux for stellar populations
dominated by K3V or earlier-type stars, which is appropriate for the
majority of objects observed in the three closest galaxies.  However,
one cluster in NGC~5253 is of particular interest, cluster 5 from
Calzetti et~al.\ (1997).  This is a highly extincted, very young
cluster, which emits strong thermal radio emission, but is also seen
in optical/ultraviolet images. Note that the exact photometry for this 
single object does not affect our overall conclusions.

\subsection{Photometry}

Aperture photometry was obtained using the PHOT task in DAOPHOT
(Stetson  1987).  It has been established that aperture corrections in
different \emph{HST} WFPC2 filters show very little difference with object
size (Holtzman et~al.\ 1996; Larsen 1999).  Thus, we used a small
aperture radius ($r=3$ pixels) to determine object colors, in order to
minimize contamination due to neighboring sources and the impact of
uncertainties in the background determination when fainter object
light is included.  However, there is a significant fraction of light
outside this radius, which varies based on how extended an object is,
and which directly impacts the total magnitude measurements.  In order
to measure aperture corrections for our sources, we first measured the 
sizes of all detected objects using Larsen's (1999) ISHAPE routine.  
Details of size measurements are given below.  
Since the sources have a large range of sizes, which leads to very
different aperture corrections, we linked the aperture corrections to
the measured size for each object.  We used the previously measured
aperture corrections from Larsen \& Brodie (2000) for ISHAPE KING30
models (for $\Delta m_{3->5}$ and $\Delta m_{5->30}$), which are
compiled in their Table~1.  These values are given for discrete values
of the measured FWHM in pixels.  We fit a second order polynomial to
these values, and used this equation to calculate aperture corrections
for each object individually.

CTE corrections were performed as described in Dolphin
(2000).\footnote{see http://www.noao.edu/staff/dolphin/wfpc2\_calib/
for updated calibrated information.}  The corrected instrumental
magnitudes were converted to standard Johnson-Cousins $U$, $V$, and
$I$ magnitudes.  Using Equation~8 and Table~7 of Holtzman et~al.\ 
(1995), the magnitudes were derived iteratively using WFPC2
observations in two filters.

\subsection{Size Measurements}

Intrinsic sizes for our entire object sample were measured using the
ISHAPE routine.  A detailed description of the code is given in Larsen
(1999), along with the results of extensive performance testing.
Essentially, ISHAPE measures intrinsic object sizes by adopting an
analytic model of the source and convolving this model with a
(user-supplied) point spread function (PSF), and then adjusting the
shape parameters until the best match is obtained.  King model
profiles with concentration parameters of $c=30$ were convolved with a
PSF, and fit individually to each object.  ISHAPE estimates the FWHM
of each cluster (in pixels), which was then converted to the
half-light (effective) radius, $r_{eff}$, as described in the ISHAPE
manual.  The input PSF to this algorithm is crucial.  We created a PSF
by hand-selecting stars in the image, and then compared the results
with those of a subsampled TinyTim PSF (when the TinyTim PSF was used,
convolution with the WFPC2 diffusion kernel was implemented).  We
found that the size estimates from ISHAPE using these two PSFs
differed by less than 20\%.  Final measurements were made using the
TinyTim PSF, since this is easily reproducible.  A single PSF was
generated for the PC CCD, and one for the WF CCDs.  We assume that
objects with $\Delta_{\mbox{FWHM}}$ 
$(\mbox{FWHM}_{measured}-\mbox{FWHM}_{PSF})$ of 0.2~pixels or 
less are unresolved, point sources.

\section{ANALYSIS}

\subsection{Deriving Stellar Properties}

One way to assess (any) differences between clusters and the field in
local starbursts is to quantify the stellar content of each.  Is the
field lacking in massive stars relative to neighboring clusters?  The
answer to this question can be used as a starting point to study the
composition of galaxies, and make inferences concerning the properties
of star formation.  One of the main advantages of using UV
spectroscopy to assess the differences between stellar content in
clusters and the field is the short timescales over which massive
stars evolve.  To illustrate the diagnostic power of UV spectroscopy
in assessing massive star content, in Figure~3 we present example
spectra from various STARBURST99 models which clearly show the effect
massive stars have on composite FUV spectra of stellar populations,
particularly on the N~IV~$\lambda1240$, Si~IV~$\lambda1400$ and
C~IV~$\lambda1550$ broad P~Cygni profiles.

In order to make a quantitative comparison between the stellar
populations dominating clusters and field regions, we created
composite ``cluster'' spectra, which are the unweighted sum of
multiple clusters.  When there were a number of clusters available, we
combined them by luminosity, making an arbitrary separation between
bright and faint clusters, depending on specifics of cluster
characteristics (luminosities) within a given galaxy.  In general,
since younger clusters tend to be more luminous than older ones of
comparable mass, this will roughly divide clusters by age.  To derive
quantitative cluster parameters, we then compare dereddened (reddening
discussed below) cluster spectra to a grid of instantaneous burst 
STARBURST99 models (Leitherer et~al.\ 1999), and recorded the results 
in Table~\ref{table:ages}. Instantaneous burst models are appropriate 
for compact clusters, since they are coeval stellar populations with small 
crossing times ($< 0.1$~Myr) compared with the models.  While 
instantaneous burst models are not an ideal physical representation of 
field regions extending over hundred of parsecs, they provide a 
consistent way to compare the stellar content of clusters and the field.  
Because we are working in the UV, the field spectra, even though they 
probably contain a mix of ages, will be dominated by signatures of the 
youngest stellar populations.  In $\S5.5.2$ we compare our field regions 
with continuous star formation models.

STARBURST99 models have been optimized to
reproduce many spectrophotometric properties of galaxies with active
star formation.  Details of the input stellar parameters to
STARBURST99 can be found in Leitherer et~al.\ (1999); here we briefly
summarize the model parameters used in this work.
The STARBURST99 UV spectral library is available for 2~metallicities:
solar and LMC/SMC ($\sim1/4$ solar).  In general, we used the
metallicity model which provides the closest match to a given galaxy's
abundance (tabulated in paper~1).  However, for some galaxies
the abundance is in-between the two models.  For these, we compared
the results from both abundance models, and adopted the one which
resulted in the better fit (lower $\chi^2$).  In general, age
estimates for individual clusters are in good agreement between the
two models.

For the instantaneous burst models, we adopted a standard Salpeter
(1955) initial mass function (IMF), with lower and upper masses of
$1~M_{\odot}$ and $100~M_{\odot}$ respectively.  Arguments in favor of
a Salpeter IMF are summarized in Leitherer (1998), and a discussion of
various studies available in the literature can be found in Tremonti
et~al.\ (2001).  High mass loss rates, as required by stellar evolution
were assumed (Meynet et~al.\ 1994).

In order to determine intrinsic properties for a stellar population,
the one dimensional spectra must be corrected for the effect of
(foreground) Galactic extinction, as well as for the dust obscuration
intrinsic to the starburst itself.  We assume foreground extinction
values from the Schlegel et~al.\ (1998) maps, and adopt the extinction
curve of Fitzpatrick (1999).  To determine the reddening internal to
the starburst itself, we compared the observed continuum of each
cluster and field spectrum with stellar evolutionary models.  The
intrinsic FUV spectral distribution of young, unobscured single
stellar populations follows a power law, with effective spectral index
$\beta$, where $F_{\lambda} \propto \lambda^{\beta}$.  Any deviation
of the power law exponent from that predicted by theoretical models is
assumed to be due to the effects of dust.  The intrinsic UV spectral
energy distributions of any young ($<10$ Myr) starburst population is
typically near $\beta \sim-2.6$.  The fit for $\beta$ was performed
over the spectral region 1240--1600~\AA.  To account for the spectral
features (which are mostly in absorption), we performed the fit
iteratively with rejection thresholds set at $2\sigma$ for the lower
bound and a $3.5\sigma$ upper bound.

We determined which model-age provided the best fit to our clusters
and field spectra by comparing the (foreground+intrinsic) extinction
corrected object spectra with instantaneous burst models from
STARBURST99 with ages from 1--100~Myr in steps of 1~Myr.  Model
fitting closely follows the techniques developed by Tremonti et~al.\ 
(2001), and the reader is referred to that work for details.
Briefly, in order to maximize the sensitivity of $\chi^2$ to spectral
regions most sensitive to the age and stellar content of the
starburst, each pixel in the spectrum is classified as belonging to
either the continuum, an interstellar line, or a stellar wind line
(based on the detailed line analysis of de Mello et~al.\ 2000).  These
were then assigned weights, with interstellar lines being given a
weight of zero, continuum a weight of one, and stellar wind lines a
weight of ten.  Because nearly all of the features which are sensitive
diagnostics of the age of a stellar population also have some
interstellar contamination, the interstellar core of the N~V~$\lambda
1240$, Si~IV~$\lambda 1400$, and C~IV~$\lambda 1550$ lines were also
masked out.  This effectively eliminates the ISM from consideration,
and relies only on stellar signatures to derive their properties.  The
routine returns the best fit age (minimum $\chi^2)$, which is given in
Table~\ref{table:ages}.

To quantify errors associated with our best fit models, we utilized
the bootstrap technique.  The residuals of the best fit were randomly
resampled and added to the model spectrum, and then rerun through the
automated fitting routine 1000 times.  The error bars associated with
the 90\% confidence interval were derived from a histogram of the
derived ages.  The age uncertainties are also given in Table~\ref{table:ages}.
For clusters, ages can be derived to within $\pm1$~Myr,
highlighting the strong effect that massive, short-lived stars
have on the UV spectrum.

\subsection{Summary of Model Fitting by Galaxy}

Figure~4 shows composite cluster and field spectra for each galaxy in
our sample.  The results from comparison with instantaneous burst
models are given in Table~\ref{table:ages}, and discussed generally
below.  Note that the figures, analysis, and discussion are based on
our first definition of the field, where we have included all regions
within $3\sigma$ of the local background, since this is a more
conservative approach and does not include obvious star clusters.
However, we have verified that the results presented here do not
change substantially for the second field definition in each galaxy.

{\it Mkn33:} Figure~4 shows the composite cluster and field spectra
for Mkn33.  The field has weak signatures of massive O~stars, and has
a best fit age of 6~Myr when compared with instantaneous burst models.
This is slightly older than found for the composite cluster spectrum.
In general, we found that separating clusters into a bright and
faint category kept any individual cluster from severely biasing the
luminosity weighted age of the composite cluster spectra.

{\it He~2-10:} Studied in a previous paper (Chandar et~al.\ 2003), the
main starburst region (A) shows individual compact clusters on an
elevated background at UV-optical wavelengths.  The nature of this
diffuse field light was investigated, and found to include a
significant population of massive stars.  We concluded that a large
number of lower mass compact clusters or diffuse SOBAs could be
responsible. The tails of this diffuse emission extend out
$\sim$100~pc from starburst~A.  The diffuse field light is fainter,
but shows the same spectral signature as that in starburst region~A.
Therefore, we see the signature of massive stars forming in the
``field'' in this actively starbursting galaxy.

{\it NGC~1741:} Figure~4 shows a sequence of three ``composite''
stellar populations.  The top panel shows the most luminous cluster in
our slit, cluster~1, the second panel is an average of fainter
clusters~2, 3, and~4, while the bottom panel shows the field spectrum.
There is a clear gradient in the stellar populations in this galaxy,
as traced by the strength of the stellar wind lines---these are
strongest in luminous cluster~1, weaker in clusters~2, 3, 4 and weaker
still in the field.  These observations suggest an age sequence, which
is quantified in Table~\ref{table:ages}.

Cluster 1 is quite young, with very strong \NV and \SiIV lines it has
a best fit age of 3~Myr.  The other three clusters all have ages of
6~Myr.  The field has the oldest composite age of 7--8~Myr, depending
slightly on which metallicity model is adopted.  Regardless, there is
no obvious signature from the most massive stars in the field region.

{\it NGC~3125:} 
The average spectrum of the two most luminous clusters shows
significantly stronger wind lines (in particular \NV and \SiIV)
relative to the field.  Cluster NGC~3125-1 also shows strong WR signatures,
including the N~IV] 1487~\AA, N~IV 1718~\AA\ lines, and the
He~II~1640~\AA\ line typically found in WN stars (see Chandar et~al.\ 2004 
for a detailed derivation of the properties for this unusual
cluster).  Fits to STARBURST99 instantaneous burst models give mean ages
of 3~Myr for both clusters, and 8~Myr for the field.

{\it NGC~3310:} The five brightest clusters have a mean age
of 5~Myr, and five faint ones have a mean age of 7~Myr.
The field has a best fit age of 8~Myr.

{\it NGC~4214:} The brightest cluster 
has strong wind lines and a young age (4~Myr).  When we average the
individual fainter cluster spectra, we
find that these objects are not only fainter, but they are also older
on average, with a mean age of 7~Myr.  Our best fit to the field gives
an age of 8~Myr.

{\it NGC~4449:} The mean cluster spectrum shows P~Cygni profiles, and
has a formal age of 5~Myr.  The field shows little evidence for the
presence of massive stars, and has a best fit age of 7~Myr.

{\it NGC~4670:} The most luminous cluster, NGC~4670-1 is 7~Myr old.
Three fainter clusters (2,3,4) are coeval, with ages of 5~Myr.  Note
also that these fainter clusters are located very close to one
another, while cluster~1 is located slightly further away. The
composite field shows the weak signature of massive stars, and has a
mean age of 6~Myr old when compared with instantaneous burst models.

{\it NGC~5253:} Tremonti et~al.\ (2001) find a difference in the mean ages
of the composite cluster and field spectra.  
Re-doing the extraction independently, we find a mean age for the nine
extracted clusters of 4~Myr and a mean age for the field regions of 7~Myr, 
similar to the earlier results.

{\it NGC~5996:} The brightest cluster is also the youngest, at 5~Myr.
The strength of the wind lines (particularly the \SiIV~feature) in the
composite faint cluster spectrum and the field establishes that
massive stars reside in both the faint clusters and the field,
even though these appear to be somewhat older than the brightest
cluster.  \NV and \CIV are slightly more pronounced in the
faint clusters than the field, so these may contain somewhat more
massive stars than the field.

{\it NGC~7552:} The two bright clusters have a mean age of 5~Myr,
while the field appears slightly older, at 6~Myr.

{\it TOL1924-416:} The brightest cluster has an age of 1~Myr, while
the youngest stellar population contributing to the UV field light has
an age of 7~Myr.

\subsection{Radial Profiles}

Radial profiles for objects in nearby galaxies can be used to
determine whether an object is extended or not.  If an object is
resolved, it is likely a cluster or group of stars; however unresolved
objects can be either stars, chance superpositions, surface brightness
fluctuations, or compact clusters.  Although size estimates are
typically made from imaging, here we attempt to also use the
information along the spatial direction of the slit to determine
whether clusters and bumps/wiggles in the field are resolved, and from
there reach some conclusion concerning the nature of the diffuse UV
light.  This is particularly valuable since STIS pixels sample even
smaller scales ($0.025\arcsec~\mbox{pix}^{-1}$) than the PC
($0.0455\arcsec~\mbox{pix}^{-1}$) CCD on WFPC2.

To measure the FWHM of bumps/wiggles in the field, we compare with the
FWHM measurement of a bright star (GD71) observed with the same
grating and slit combination.  We measure the FWHM of star GD71 to be
$\sim0.08\arcsec$, and take this to represent a point source.  Because
the size estimates for the faint peaks in the field are likely not
very accurate, we only determine whether these appear resolved or not,
relative to the measurement of GD71.  We discuss the results of this
exercise in $\S5.4$.

\section{DISCUSSION}

\subsection{Is Scattered UV Light Responsible for the Field?}

Could the scattering of UV photons explain the presence of diffuse UV
emission in local starburst galaxies?  In this scenario, UV photons
originate in young stars embedded in the massive clusters.  Since most
of the non-ionizing UV radiation produced in star forming regions
escapes the local cluster (e.g. Misiriotis et~al.\ 2001; Hippelein et~al.\ 
2003), it is free to be scattered by interstellar dust.  Recent
GALEX UV images of nearby spiral galaxies clearly show the presence of
diffuse UV emission between spiral arms (e.g., M101; Popescu et~al.\ 
2005) and at large galactocentric radii (e.g., M33, M31, and M83;
Thilker et~al.\ 2005a,b).  When combined with observations at longer
wavelengths, the spectral energy distributions (SEDs) of spirals are
well reproduced by galaxy components which include a layer of
diffusely distributed cold dust (e.g., Popescu et~al.\ 2000).  Far-IR
and submillimeter imaging of nearby spirals have directly shown that
this layer of cold dust covers the entire extent of the disk
(e.g. Haas et~al.\ 1998; Hippelein et~al.\ 2003; Popescu et~al.\ 2005).
In M101, a comparison of GALEX UV and ISO far-IR images showed that
the ratio of UV/far-IR emission varies between arm and inter-arm
regions (Popescu et~al.\ 2005).  These differences are
consistent with the inter-arm UV emission being scattered light
originating from massive stars in the spiral arms.  Similarly, the
diffuse far-UV emission observed beyond the $\mbox{H}\alpha$ disk in
M33, M31 and M83 (Thilker et~al.\ 2005a,b) is interpreted as
non-ionising UV radiation escaping from the outermost H~II regions and
then scattering off dust grains further out in the disk.  

Could the diffuse UV light observed in local starburst galaxies result
from such scattering of UV photons?  Scattering by dust can change the
original spectral shape (i.e., make the spectrum appear bluer), but
cannot add or subtract narrow spectral features, such as P~Cygni
profiles.  In this work, we have shown that the young, massive stars
which dominate the UV flux from clusters have an O-type spectrum, with
prominent P-Cygni profiles.  By contrast, the UV light originating in
the field shows a B-type stellar spectrum.  Because the spectra from
the two environments have {\it different\/} stellar features, massive
stars powering the luminous clusters are not the original source for
any UV photons scattering off dust grains.  However, if lower mass,
B-star dominated clusters are found in the field, we cannot rule out
the possibility that photons leaking out from these lower mass
clusters are scattered, and thus produce the field UV light.
Regardless, the differences in spectral features observed between the
field and clusters demonstrates that the UV emission in each environment
originates in a {\it different stellar population\/}.

How can our evidence that the diffuse UV light in starburst originates
in a (faint) stellar population be reconciled with the claim that the
diffuse inter-arm UV light in M101 is due to scattering off dust
grains?  First, the geometry and distribution of dust in large spirals
may differ substantially from that in dwarf systems such as those
which often host starbursts.  Second, Popescu et~al.\ (2005) point out
that even in the most extreme interarm regions discrete sources of UV
emission are observed in full-resolution GALEX images.  Given the
low-resolution of the GALEX images ($6\arcsec$~FWHM) with respect to
\emph{HST}, it is plausible that some fraction of the diffuse inter-arm
emission in M101 arises from a dispersed stellar population rather
than from scattering due to dust.  For example, the GALEX images are
not sufficiently deep (and do not have the spatial resolution) to
detect individual B~stars (which are a plausible contributor to the
diffuse UV field light).  Rather, these observations are appropriate
for studying stellar complexes which have sizes of many tens to
hundreds of parsecs at the distance of M101.  An analysis of WFPC2
images for NGC~4214, NGC~4449, and NGC~5253 clearly reveal a
population of dispersed blue point sources, which are consistent with
a population of B-supergiants ($\S5.3$).  We suggest that deep UV
imaging of interarm regions in spirals, with sufficient depth to
reveal the presence of individual B~stars
could resolve whether the diffuse UV emission in M101
arises from a diffuse layer of dust, or from a discrete stellar population.

\subsection{General Summary of the Stellar Content in Field Regions of Starbursts}

\subsubsection{FUV slope measurements}

In Table~\ref{table:beta} we compile slopes, $\beta$, measured for the
composite field and cluster spectra after correction for foreground
extinction, but with no correction for extinction due to the host
galaxy.  Details of the $\beta$ measurements were provided in $\S4.1$.
The main uncertainties in the values of $\beta$ come from the fact
that a power law does not always provide a good fit to the data,
mostly due to line blanketing in the spectra (although we have
excluded the most severely line blanketed spectral regions longward of
1640~\AA).  

In general, the intrinsic slope of a stellar population becomes redder
with age.  However, a comparison of the slopes for single stellar
populations younger than $\sim$10~Myr (typical ages of clusters
observed in our starburst galaxies), with those for continuous
formation models which fit our data show that both have similar
values.  For example, a continuous star formation model with Salpeter
IMF and upper mass cutoff of $30~M_{\odot}$ has an intrinsic slope,
$\beta=-2.5$ (see $\S5.5.2$ for a discussion of continuous star
formation models and their comparison with the data).  Because the
dominant/youngest UV populations in the clusters are somewhat younger
than those in the field (as discussed in detail below), one might
expect that the intrinsic slope for the clusters is somewhat bluer
than for the field.

In Table~\ref{table:beta} however, $\beta$ measurements for the field
appear comparable to or bluer (steeper) for half (6/12) of the field
spectra when compared with the composite cluster spectra in the same
galaxy.  Because the spectral features in the UV allow us to precisely
age-date the stellar populations in the clusters and the field, we
know that the dominant stars in the field are generally older or less
massive than those in the clusters.  Therefore one might expect that
$\beta$ for the clusters would be bluer than for the field.  However,
our study establishes that the field is not {\it categorically\/} redder
than young clusters in starbursts.  A more likely explanation for the
observed difference in slopes between clusters and the field is
extinction variations between stellar populations residing in each
environment, with the field showing less reddening.  Because the
UV-bright clusters are quite young, it is likely that they
still have at least some residual natal ISM and dust surrounding them,
leading to redder slope measurements.

The bluer, less reddened slopes found in the field regions of some local
starbursts when compared with star clusters have implications for the
overall extinction estimates derived for high redshift galaxies.
This is particularly important when UV spectra of whole high-z
galaxies are compared with narrow-slit or small-aperture UV spectra of
local star-forming galaxies. Inferences of differences in stellar
population compositions and/or extinction characteristics may stems
from the fact that the UV spectrum of the local galaxy is only
targeting one component (e.g., one cluster) of the entire star forming
region.

\subsubsection{Characterizing the Field Stellar Population}

We find that the overall stellar content in our observed STIS field
regions of nearby starburst galaxies can be divided into three broad
categories: the clear presence of O~stars, a clear lack of O~stars,
and an intermediate class showing a weak signature of O~stars.
He~2-10 is unique to our sample, with the youngest mean field stellar
population of any sample galaxy, clearly showing the signature of 
O~stars.  The other extreme includes 7~galaxies, which show field
spectra lacking in the most massive stars (absent P~Cygni profiles),
similar to that found by Tremonti et~al. (2001) for NGC~5253.  This
category also includes NGC~1741, NGC~3125, NGC~3310, NGC~4214,
NGC~4449, and Tol1924-416.  Table~\ref{table:ages} shows that these
have best fit ages of $\geq7$~Myr, and inspection of the field spectra
in Figure~4 confirms the lack of massive stars.  The third
category comprises Mkn33, NGC~4670, NGC~5996, and NGC~7552, which have
mean field ages of 6~Myr, and show weak signatures of O~stars.

The closest galaxies (i.e.\ those within 4~Mpc; NGC~5253, NGC~4214, and
NGC~4449), where we have the best spatial resolution and the easiest
time distinguishing low level clusters from the field, all clearly
lack broad line profiles of N~V~$\lambda1240$, Si~IV~$\lambda1400$,
and C~IV~$\lambda1550$ that are characteristic of O~star winds, in
stark contrast to the composite cluster populations in each galaxy.
This result holds for both field definitions, whether we are excluding
all peaks above $\gea3\sigma$, or only excluding targeted, luminous
star clusters.  Therefore we are not concerned that our definition has
{\it a priori\/} excluded O~stars from the extracted field spectra.  Not
surprisingly, we see more variation in the field regions of more
distant galaxies.  Some, e.g., NGC~3125, NGC~3310, NGC~1741, and
Tol1924-416 (all further than 10~Mpc), clearly lack the signature of
massive stars, and are very similar to the field spectra for the
closest galaxies.  Others, such as Mkn33, NGC~4670, NGC~5996, and
NGC~7552, again more distant than 10~Mpc, show weak signatures of 
O~stars.  For galaxies in this latter group, our field extraction likely
includes some young, relatively massive clusters, which provide the
weak O~star signature.  This bias with distance is not unexpected;
however, it is gratifying that a number of the more distant galaxies
in our sample show a result similar to the closest targets.  We
conclude that the field regions of local starbursts (except He~2-10)
are not conducive to the formation of isolated massive stars (as
suggested is occurring in the nuclear regions of M51; Lamers et~al.\ 
2002), in the random field regions covered by our longslit pointings.

For galaxies with mean derived field ages of $\geq7$~Myr, we find
that the faint clusters are only 1--3~Myr younger than the youngest
stellar populations contributing to the field regions. Implications for 
these relatively small differences in age between the dominant stellar 
populations in clusters and the field are discussed in $\S5.5.4$.  

Based on the results described above, He~2-10 appears to be unique in
our sample.  Because the dominant stellar population in the field
formed coevally with the nearby clusters, we estimated the UV
contribution of the compact clusters in starburst region A to the
total UV flux.  Assuming that the clusters which did not fall in our slit
have the same age as those which were studied, we find that 40\% of
the light comes from the clusters, leaving the remaining 60\%
originating in the field.  This is still a lower limit since some
clusters may have dissolved, but provides a firm lower limit to the
amount of UV light coming from clusters in this starburst.

Our main result from analysis of the stellar content of clusters and
the field regions of starbursts is that the field generally shows
weaker or absent P~Cygni profiles compared with clusters.  Thus the
fraction of high mass stars that are visible is smaller in the field
than in the bright clusters.  Here we have simply showcased this
result by presenting the best fit ages derived by comparing
instantaneous burst evolutionary synthesis models with our extracted
spectra.  Although this is not the best physical representation of the
field, it is a simple exercise which illustrates the difference in
massive star content.  This result in starbursts is similar to that
found for O~stars in the Galaxy and LMC.  Van den Bergh (2004) used a
catalog of Galactic O-type stars (Maiz-Apellaniz et~al.\ 2004) and 
found that O-type stars in clusters and associations
have earlier types (presumarly larger masses or younger ages) than
those in the general field.  In the LMC, Massey et~al.\ (1995) finds
that although massive stars are born in the field, they form more
rarely in this environment than in clusters and associations.

Could our observed lack of massive field stars result from statistics,
due to the relatively small areas covered by our STIS slit, or
possibly due to modest star formation rates?  In order to assess this
possibility, we use our extracted spectra to estimate the mean star
formation rate ($\overline{SFR}$) implied for the field, and also to
predict the number of O stars expected to reside in the projected area
covered by our field regions assuming a normal Salpeter IMF with an
upper stellar mass limit of $100~M_{\odot}$.  To calculate the
$\overline{SFR}$ in the field regions of local starbursts, we compare
our dereddened (for both foreground and intrinsic extinction) field
luminosity at 1500~\AA\ (corrected for distance) with that predicted by
a continuous STARBURST99 model with Salpeter IMF.  The parameters used
for this calculation, along with the derived field $\overline{SFR}$,
are compiled in Table~\ref{table:sfr}.  Meurer et~al.\ (1995) used UV
FOC images to estimate the star formation rate for a number of local
starbursts.  If we correct their derived $\overline{SFR}$ for massive
stars by a factor of 2.16 to extend the IMF down to $1~M_{\odot}$, we
find values within a factor $\sim$2--3 for the five galaxies in common
between the two works (NGC~3310, NGC~4670, NGC~5253, NGC~7552, and
Tol~1924$-$416); a high value for the $\overline{SFR}$ in NGC~7552 is
found in both works.

To assess whether sampling statistics is responsible for missing
field O~stars, we assume a continuous star formation episode from the
STARBURST99 models, appropriate for the spatially extended field
regions, with Salpeter IMF, $M_{low}=1~M_{\odot}$ and
$M_{up}=100~M_{\odot}$, and with a metallicity appropriate for each
galaxy.  The number of O~stars in this model equilibrates after
10~Myr, and we scale the star formation rate (as calculated above) and
projected area of the field regions to estimate the predicted number
of O~stars under these assumptions.  The results are tabulated in
column~6 of Table~\ref{table:sfr}.  This exercise establishes that our
spectra cover large enough portions of the field that, if the field
IMF is the same as that for clusters, and massive stars in the field
are not heavily enshrouded in dust, their signature should be clearly
visible in our UV spectra.

\subsection{Photometry Results for NGC~4214, NGC~4449, and NGC~5253 and
Implications}

In order to give a broader and complementary picture of the stellar
content in the field regions of the closest galaxies, here we analyse
photometry of the archival WFPC2 multiband imaging.  This is not meant
to be a rigorous examination of the stellar content.  Rather, we are
interested in obtaining additional information which will shed light
on the nature of the field populations observed in our STIS spectra.

In Figure~5 we plot two-color and color-magnitude diagrams for
sources detected in the WFPC2 CCD where our STIS slit location falls.
The top panel in each figure shows the $U-V$ vs.\ $V-I$ color-color
diagram; the object colors have been dereddened by the foreground
extinction values of Schlegel et~al.\ (1998).  Resolved
($\Delta_{\mbox{FWHM}}>0.2$~pix) and unresolved
($\Delta_{\mbox{FWHM}}\leq0.2$~pix) objects are shown in different
symbols.  We also show the stellar evolutionary model predictions of
Bruzual \& Charlot (2003) for star clusters as a function of age, with
three separate metallicity models.  Although local starburst galaxies
tend to have LMC/SMC-type metallicity, the tracks for these lower
metallicities poorly reproduced observed colors of red giant stars
(Massey \& Olsen 2003).  The agreement between models and observations
are better at solar metallicity, and therefore we include cluster
tracks at this metallicity.  The bottom panel in each figure plots
$M_V$ versus the $V-I$ color.  None of the data have been corrected
for any internal extinction.

For each galaxy, NGC~4214, NGC~4449, and NGC~5253, the resolved and
unresolved points fall in the same parameter space.  In general, the
$U-V$ vs.\ $V-I$ color-color diagrams of resolved objects detected in
NGC~4214, NGC~4449, and NGC~5253 show a mostly young stellar
population ($\lea10$~Myr) when compared with stellar evolutionary
models of star clusters.  These resolved objects continue down to
relatively faint magnitudes.  Inspection of the WFPC2 images suggests
that some of the objects which have $\Delta_{\mbox{FWHM}}$
measurements greater than 0.2~pixels are actually blends of two or
more stars.  However, the majority of resolved objects appear to be
individual sources.  These resolved objects, which appear to be good
star clusters plus a handful of faint background galaxies, have
luminosities extending down to at least $M_V=-5$, despite their very
blue colors.  We also see from the location in the WFPC2 images that
individual, blue point sources are dispersed throughout each starburst
galaxy.  The very blue colors suggest that these are young, massive
stars.  However, without spectroscopy, the degeneracy in optical
colors makes it difficult to determine whether these are O or B~stars.
Based on the field signature found in our STIS spectroscopy, at least
in these three galaxies, the dispersed population of blue point
sources cannot be O~stars.  We conclude that the combination of
photometry and spectroscopy supports the presence of a dispersed
population of B~stars in each starburst.

In the CMDs plotted in the lower panels, we don't see evidence for
differences in the absolute luminosities of resolved and unresolved
sources.  In all three galaxies, we find unresolved sources which have
absolute V~magnitudes brighter than $-8$, which is relatively rare for
individual stars.  Due to their high luminosity, these may actually be
superpositions, groups, or very compact clusters, and not individual
stars.

Because it appears likely that any individual stars in the WFPC2
images are B rather than O~stars, which have significantly longer
lifetimes and could easily have formed in clusters that have since
dissolved, the data do not require on-going star formation in the
field, as suggested for NGC~1705 by Tosi et~al.\ (2001) and Annibali 
et~al.\ (2003).  In NGC~1705, these authors found, via comparison of the
data with synthetic color-magnitude diagram techniques, that to
account for the bluest stars in the observed CMD, required a burst of
star formation $\sim$3~Myr ago.  If this scenario is accurate, it
implies that individual massive stars are forming in the field regions
of NGC~1705, and not in stellar clusters.  This conclusion differs
from ours, since we find no evidence for very young O~stars forming in
the field, and it is possible that the dispersed B~star population
originally formed in clusters more than 10~Myr ago.  Can these
different conclusions be reconciled?  One possibility, is that a
number of the unresolved, blue objects in NGC~1705 are actually young,
low mass star clusters.  In NGC~4214, NGC~4449, and NGC~5253 we find
evidence for resolved, and hence likely star clusters, with blue
colors down to faint magnitudes.  NGC~1705 is even further away than
these three galaxies, with a distance modulus $(m-M)_0$ of
$28.54\pm0.26$, making it more difficult to sort out compact clusters
from individual stars based on size measurements.  Higher angular
resolution studies of NGC~1705 could settle this issue.

\subsection{Spatial Profile Analysis and Implications for the Field Stellar Population}

In this section, we use our crude fits to low level peaks in the
spatial cut to glean additional information concerning the nature of
the diffuse light.  Because the contribution to the field is subject
to biases with galaxy distance, where relatively higher mass clusters
may be ``hiding'' in the field of more distant galaxies, we again
focus on the closest galaxies in our sample: NGC~4214, NGC~4449, and
NGC~5253 all located within 4~Mpc.  For local peaks within the
background which can be reasonably well isolated and don't show
obvious signs of being blends, we determined whether they were
resolved or unresolved (precise size estimates at these low S/N values
would have large uncertainties).  These three galaxies show fractions
of {\it resolved\/} low level peaks in the field (typically we found
$\sim$13--17 such peaks per galaxy) from $\sim30$\% to 60\%.  Based on
our measurement of the spatial profile of the star GD71, unresolved
objects have a FWHM $\sim0.08\arcsec$.  In the STIS spatial cuts, we
consider a peak to be resolved if it has a FWHM measurement of
$0.09\arcsec$ or greater.  At the distances of NGC~4214, NGC~4449, and
NGC~5253, this means that compact clusters with FWHM measurements
smaller than $\sim0.2$~pc would be considered unresolved.  Note that
when compared with the brighter peaks which we have defined as
clusters, a smaller relative fraction of the faint peaks in the field
are resolved within a given galaxy.

Based on the field spectral signature, we know that these are not
individual O~stars, although they are precisely in the flux range
where we would expect to see individual O~stars included in our
definition of the field.  Could we be seeing the continua from a
dispersed B~star population?  To check if the observations are deep
enough to detect the continua from individual B-stars, we first
estimated the approximate detection limit for continua in NGC~4214
from the STIS spectra (few$\times10^{-16}$~$\mbox{erg~s}^{-1}$ at
1500~\AA), NGC~4449 (few $\times10^{-16}$~$\mbox{erg~s}^{-1}$), and
NGC~5253 ($10^{-16}$~$\mbox{erg~s}^{-1}$) in our field regions.  We
then estimated the (unreddened) flux expected for a B0 supergiant at
1500~\AA, assuming a temperature of 30,000~K, and a mass of
$17.5~M_{\odot}$ from Kurucz (1993) model atmospheres.  We find that
in the most optimistic scenario where a single B0 supergiant has no
reddening, the flux would be just above our detection limit.  However,
this scenario is not very likely, since lower mass B~stars are much
more numerous than B0 stars, and because it is unlikely that such
stars have no reddening, as demonstrated for example, by a comparison
of spectral type and observed color for B~stars in the Humphreys
(1978) catalog.

Therefore, we conclude that while some
of the weaker unresolved peaks in our field spectra could be the
continua from individual early B~supergiants, it is unlikely that all
of the unresolved peaks come from individual B~stars.  This does not
however preclude a population of dispersed B~stars from contributing to the
continuum level observed in our UV spectroscopy of the field.

Because our calculations suggest that individual B~stars are not
responsible for all of the individual faint peaks seen in the field
portions of our STIS spectra, it is likely that at least some of the
unresolved peaks which we observe in the field regions of these local
starbursts are either chance superpositions of B~stars, or compact
star clusters.  The question then becomes are these young, low mass
clusters, or older more massive but faded clusters.

We can also use our photometry to help place limits on the properties
of star clusters contributing to our STIS spectra, since it is
possible to match individual sources seen in archival $U$~band WFPC2
images with the objects in our slit.  If we restrict the object
samples in the WFPC2 images to those within 50~pc of the STIS slit, we
set approximate detection limits $M_V=-5.8, -5.3, -5.9$ for NGC~4214,
NGC~4449, and NGC~5253 respectively, from available WFPC2 imaging.  A
comparison of the $U-V$ vs.\ $V-I$ colors of resolved objects to the
stellar evolutionary models of Bruzual \& Charlot (2003) suggests that
these sources are almost exclusively young, with likely ages of
several Myr (although ages derived from integrated photometry,
particularly for such low mass clusters, are subject to relatively
large uncertainties).  By scaling the luminosity of a 7~Myr star
cluster of $10^6~M_{\odot}$ ($M_V=-15.5$ according to STARBURST99
models), we find that any young clusters which are not observed in the
archival WFPC2 images would have masses of several hundred solar
masses and less (i.e. $\lea\sim300-500~M_{\odot}$).  We note that this
is similar to the lowest mass star cluster detected by Tremonti et~al.\ 
(2001) in their study of NGC~5253 using STIS long-slit
spectroscopy.  Older clusters at similar luminosities could also
contribute to the field, and would be more massive.  In $\S$ 5.5.3 and
5.5.4, we explore in turn the possibilities that young, low mass
clusters and older, dissolving/fading clusters contribute to our
observed field regions.

\subsection{Scenarios for Understanding the Field Stellar Population}

Our main result is that the diffuse UV light in starbursts lacks the
strong O-star wind features observed in neighboring UV-bright star
clusters.  In this section, we explore four possible scenarios which
can explain this result:

\begin{itemize}

\item Star formation occurs in both clusters and the field, but the
most massive stars in the field remain in the deeply enshrouded phase
longer, and thus to do not contribute to the UV flux.

\item Star formation occurs in both the field and clusters, but there
is little or no high mass star formation occuring in the field.  This
hypothesis requires that stars formed in clusters and the field have
different IMFs.

\item Star formation occurs primarily in clusters, and what we are
calling the field is composed of young, but lower mass clusters (the
low mass end of the powerlaw cluster mass function), which do not form
O stars.

\item Star formation occurs primarily in clusters, and as clusters age
they dissolve, releasing their remaining stars into the field.

\end{itemize}

In \S5.5.1--5.5.4 below, we discuss each scenario in turn including
general implications for field formation.

\subsubsection{Are Deeply Enshrouded O Stars Hiding in the Field?}

One scenario which could explain the observed difference in massive
star signatures between clusters and the field is the possibility that
star formation occurs in both environments, but that field O~stars
remain obscured (at UV wavelengths) on timescales comparable to their
lifetimes, and thus longer than similar stars in clusters.  There are two
mechanisms which influence when a massive star becomes visible at
optical/UV wavelengths: (1)~radiation and winds from massive stars
blow out the parent molecular cloud, and (2)~OB associations tend to
drift away from their parent clouds, since stars and gas experience
different rates of friction, and thus have different velocities.  In
order to reproduce both the wind features and continuum slopes
observed in clusters and field spectra, individual O~stars formed in
the field can neither blow off their natal material nor drift away
from their parent cloud on timescales of $\sim$6--7~Myr, otherwise
their signature would be observed in our field spectra.  This
restriction would not apply to O~stars formed in clusters.  The
typical travel time, although quite uncertain, has been given as
1--3~Myr (de Jong \& Brink 1986; Leisawitz \& Hauser 1988).

The problem with this scenario is that it requires fine-tuning of both
drift and blow out timescales for field O~stars, which must in turn be
longer than those for similar stars formed in clusters.  Therefore, we
conclude that it is unlikely that deeply embedded, individual O~stars
are forming in the field regions of these starbursts, and we have just
missed them because little UV flux can escape.

\subsubsection{Constraints on Field IMF}

If star formation is occuring in both field and clusters, the lack of
massive stars observed in the field could arise if the IMF differs
from that in clusters.  We found in $\S5.2.2$ that instantaneous burst
STARBURST99 models older than 6--8~Myr produce reasonably good fits
to the field spectra.  However, these are unlikely to be realistic
representations of spatially extended field regions.  Because these
field regions cover a minimum linear distance of $\sim$100~pc, it is
unlikely that all the stars formed at the same time.  Here, we compare
continuous star formation models with the extracted field spectra and
discuss implications.  These models take $\sim$10~Myr to equilibrate,
after which the FUV spectrum changes very little as a function of age.
Therefore, we fix the age at 50~Myr, and assume a standard Salpeter
IMF (slope $\alpha=-2.35$), and lower mass cutoff
$M_{low}=1~M_{\odot}$ (the results of our UV spectroscopy are rather
insensitive to the lower mass cutoff to the IMF).  There are two
possible scenarios which can explain the lack of massive stars
observed in field regions; both assume implicitly that star formation
is occurring {\it in situ\/} in the field.  In the first set of models,
we fix the age of the continuous star formation episode at 50~Myr,
assume a Salpeter IMF, and determine whether our field spectra are
consistent with an upper mass cutoff which differs from that for
clusters.  The second set of continuous star formation models fits for
the best IMF slope, with an upper mass cutoff of $100~M_{\odot}$.

For the first set of models where we fit for the upper stellar mass
cutoff, we have recorded the best fit in column~2 of
Table~\ref{table:field}.  Qualitatively, the lack of massive stars
which is expressed by older ages when comparing with instantaneous
burst models, translates to upper mass cutoffs lower than
$100~M_{\odot}$.  The effect of varying the upper mass cutoff is shown
in the second column of plots in Figure~3.  Clearly, as the upper mass
cutoff is lowered from $100~M_{\odot}$, the signature of massive
stars, particularly the P~Cygni profiles for N~V~$\lambda1240$,
Si~IV~$\lambda1400$, and C~IV~$\lambda1550$, disappears.  In order to
reproduce the observed field spectra, the upper stellar mass cutoff in
these models has to be lowered from the $100~M_{\odot}$ used in the
instantaneous burst models, and which provides a good fit for the clusters.
For the field spectra in most target galaxies, particularly in the
nearest galaxies and those distant ones which also do not show the
signature of massive star wind lines, the preferred upper mass cutoff
is 30--$50~M_{\odot}$.

A second possibility is that the IMF of stars formed in the field is
steeper than those forming in clusters.  For example, Massey et~al.\ 
(1995) find field stars as massive as those formed in clusters and
associations in the Magellanic Clouds, even after ensuring that
runaway stars are excluded.  We tested this possibility by comparing a
continuous STARBURST99 model fixed at an age of 50~Myr, with lower and
upper mass cutoffs of $1~M_{\odot}$ and $100~M_{\odot}$ respectively,
and allowed the slope of the IMF, $\alpha$, to vary between a normal
Salpeter value of $-2.35$ and the $-5.0$ found by Massey et~al.\ (1995)
for the field in portions of the Magellanic Clouds.  The most
important influence of the massive stars is seen in the wind lines,
and so we only weighted N~V~$\lambda1240$, Si~IV~$\lambda1400$, and
C~IV~$\lambda1550$ in our fits.  In the third column of plots in
Figure~3 note in particular the rapidly decreasing
strength in the N~V~$\lambda1240$ line.  We find in general that the
preferred value of $\alpha$ is between $-3.0$ and $-3.5$.
This is steeper than Salpeter, but not as steep as found for the
Magellanic Clouds.  Comparison of our extracted field regions with
continuous star formation models rule out an IMF slope as steep as
$-4.0$ for our galaxies.  Due to the possibility that our extracted
field regions include star clusters, and hence some additional massive
stars which would result in a flatter slope, our values of $\alpha$
represent a firm lower limit to the field IMF slope.  Therefore, if
star formation is occuring {\it in situ\/} in the field, and not as a
result of dissolving or aging star clusters, the field is much less
likely to produce massive stars than the cluster environment.

\subsubsection{Constraints on the Contribution of a Young Cluster Population
to the Field}

The ultraviolet emission from the field regions surrounding star
clusters in our target starbursts, in all cases with the exception of
He~2-10, lacks the strong O-star wind lines of N~V, Si~IV, and C~IV
that are signatures of the most massive stars.  This lack of high mass
stars is not due to undersampling field regions in nearby starbursts,
as the field regions contain a considerable amount of light.  If we
reject that the IMF slope and upper stellar mass cutoff of field
regions differs from that found in clusters, there are two remaining
possibilities which can explain both the observed spectral signature
of the UV field and the strength of the faint peaks in the field
regions.  One possibility is that the field is composed of clusters
which are coeval with those observed in our slit, but have lower
masses.  The second is that the field is made up of older/dissolving
star clusters.  The difference in these two scenarios is primary one
of {\it age\/}.  Although galaxies almost certainly form clusters with
a range in both age and mass, for simplicity, we assume here that the
presumed population formed coevally with the luminous clusters.
These clusters would represent the lower extension
of the observed powerlaw cluster luminosity function (see e.g.,
Whitmore et~al.\ 2003 and references therein).  

We determined in the previous section that clusters with ages of
several Myr and masses of several hundred solar masses would not have
sufficient S/N to be classified as clusters in our spectra, even in
galaxies closer than 4~Mpc.  Furthermore, despite their youth, it is
reasonable to expect that such low mass clusters may be lacking in the
most massive O~stars, yet still be able to form B~stars.

There are two current arguments regarding how the upper mass portion
of the stellar IMF is populated in lower mass clusters.  The first
suggests that the distribution of stellar masses is a random sampling
of the initial stellar mass function, which naturally would result in
the formation of very few massive O~stars in lower mass clusters.  For
example, a recent study by Oey, King, \& Parker (2004) of the number of
OB stars per cluster or association ($N_{*}$) in the Small Magellanic
Cloud, concludes that the distribution of $N_{*}$ is consistent with
these being the most massive stars in groups of lower mass stars.
The second argument (e.g., Weidner \& Kroupa 2004) suggests that there
is a fundamental upper stellar mass limit which depends on the total
mass of a cluster.
From their Figure~4, a cluster at the detection limit of our study
would have an upper {\it stellar\/} mass limit of $\sim10~M_{\odot}$.
We performed a simple experiment to investigate whether stars with
masses $\gea20~M_{\odot}$ ($\sim$O~stars) are likely to form in
clusters with masses of a few~100$~M_{\odot}$.  We randomly drew stars
with masses between $0.1$ and $100~M_{\odot}$, from a distribution
having a Salpeter slope ($-2.35$).  Out of 1000 such simulated
clusters, roughly 200--300 had at least one star more massive than
$20~M_{\odot}$.  Therefore, if young, lower mass clusters dominate the
diffuse UV light, our limits imply that there must be a sliding upper
stellar mass limit, which is related to the total mass of the cluster,
as suggested by Weidner \& Kroupa (2004).

\subsubsection{Constraints on the Timescale of Dissolving Clusters Contribution
to the Field}

Because the field regions of starbursts contain less massive stars
than the clusters, a fourth possibility that the main difference
between the cluster and field stellar populations is truly one of age.
The field could be composed of stars which originally formed in
clusters, but which dissolved and released their remaining stars to
the field.  {\it If\/} the field is the product of dissolving star
clusters, than we can place constraints on the timescales that
clusters in these starbursts must disintegrate.  Based on the youngest
and dominant ages of the field compiled in Table~\ref{table:ages} when
compared with those for star clusters, we suggest that clusters which
contribute to the field need to be destroyed on very rapid timescales,
of order 7--10~Myr.

The evolution of cluster systems in general includes a number of
disruptive processes, such as mass loss from stellar evolution, and
stellar evaporation due to external gravitational shocks and internal
two-body relaxation (e.g., Fall \& Zhang 2001).  However, these
mechanisms occur over relatively long timescales ($\gea 10^8$~yrs)
when compared with the 7--10~Myr estimated above.  A recent study of
the age distribution of star clusters in the Antennae galaxies
suggests a very rapid decline.  Fall (2004) and Fall, Chandar, \&
Whitmore (2004) find that the number of clusters as a function of age
falls by a factor $\sim$10 by the time the cluster population has
reached an age of 10~Myr.  This rapid decline is seen for different
mass ranges above $3\times10^4~M_{\odot}$.  Number counts of embedded
clusters in the solar neighborhood with masses lower by factors
10--$10^2$ also show a similar steep decline (Lada \& Lada 2003).  Fall
et~al.\ (2004) suggest that this short disruption timescale indicates
that the majority of (but not all) star clusters in the Antennae end
up gravitationally unbound, even if the cloud from which they initially
formed was initially bound.  The ionizing radiation, stellar winds, and
supernovae explosions from massive stars could easily remove a significant
fraction of the ISM from a protocluster, leaving the stars within it
gravitationally unbound and freely expanding. If this is the case, 
such clusters would become very difficult to detect after
$\sim$10~Myr.  The very short timescales for cluster dissolution
inferred from our data in local starburst galaxies, {\it if\/} dissolving
clusters are responsible for the diffuse UV field emission, are
consistent with the timescales of cluster disruption predicted by the
free expansion model.  The advantage of using UV spectroscopy, as we
have done here, is the ability to track the presence of the most
massive stars, and directly translate these into age differences
between clusters and the field.

\subsection{High S/N Field Template}

The differences observed in the field and cluster spectra of our
galaxy sample have potentially important implications for the rest-UV
observations of galaxies at high redshift.  Because of their small
angular extent, entire galaxies are observed at high redshift, while
the best local counterparts are studied in a more piecemeal way.
Because the diffuse UV light from starbursts typically dominates the
output from stellar clusters, it is very important to include this as
an ingredient in spectral synthesis of galaxies at high redshift
(although Steidel et~al.\ 1996 show that a comparison of $z\sim3$ Lyman
Break galaxy spectra with those of individual local clusters also give
reasonable fits).

We have combined the extracted field spectra for five metal-poor
galaxies (Mkn33, NGC~4214, NGC~4449, NGC~4670, and NGC~5253) which do
not show the signature of massive stars, in order to make a high S/N
template of a low metallicity ``field'' spectrum\footnote{The template
field spectra are available from the STARBURST99 website:
http://www.tsci.edu/science/starburst} (Figure~6).  Typical abundances
for the galaxies are $\sim$1/3--1/5 solar, which is a good match to
abundances measured for high redshift galaxies (Pettini et~al.\ 2000).
This has a characteristic spectrum which is ``older'' than individual
clusters (i.e.\ lacking the signature of massive stars), and is well
represented by continuous star formation models.

\section{SUMMARY AND CONCLUSIONS}

We have used STIS long-slit FUV spectra of twelve local starburst
galaxies to study the stellar content of the diffuse, UV luminous
field regions found between prominent star clusters.  The extracted
spectra are compared with STARBURST99 stellar evolutionary synthesis
models.  He~2-10 is unique to our study, in that it contains the
signature of massive O~stars in the field regions.  The composite
field spectrum is very similar to that for four coeval clusters in our
slit.  We estimate that 40\% of the light in the far UV comes from
observed compact star clusters, providing a firm lower limit to the
amount of UV light originating in clusters in this starburst.

With the exception of He~2-10, the clusters and field regions in the
other 11 target galaxies exhibit pronounced differences.  Most of the
UV-bright clusters are quite young ($\lea6$~Myr), and show the strong
P~Cygni profiles found in O stars.  The neighboring field regions
however, clearly lack these wind features.  In particular, the nearest
galaxies (NGC~4214, NGC~4449, and NGC~5253), as well as a number of
more distant ones, have B-star dominated field spectra.

We include an analysis of \emph{UVI} WFPC2 imaging for these three closest
galaxies, in order to better understand the stellar populations which
contribute to the field.  Photometry of observed sources in all three
galaxies reveals populations of blue resolved and unresolved sources.
The colors and luminosities of resolved objects are indistinguishable
from those of unresolved sources.  This suggests the
presence of low mass, resolved star clusters, as well as a dispersed
population of blue stars.  Because we see no evidence for the presence
of O~stars in the field regions of these galaxies based on our
(limited coverage) STIS spectroscopy, we suggest that these are a
dispersed population of B~stars.

The spatial profiles along the slit show that the field regions are
not smooth, but rather contain numerous faint peaks and valleys.  An
analysis of the faint peaks in the field regions of our STIS spectra
suggests that these arise from discrete stellar populations.
Roughly 30--60\% of these peaks are resolved.  Our calculations show
however, that the STIS spectra are just sufficient to detect the
continua from unreddened, individual early B~supergiants in the field.
Therefore, while the population of dispersed B~stars discovered in the
archival WFPC2 images probably do contribute to the field spectra,
they are probably not responsible for all of the faint peaks we
observe.  We conclude that a significant fraction of these faint peaks
are likely small groups or clusters of stars, regardless of whether
they are resolved or not.

We explore four possible scenarios to explain our observation that the
field contains lower mass stars than neighboring clusters.

(1) If star formation occurs {\it in situ\/} in both clusters and
the field, and O~stars formed in the field stay in the enshrouded
phase longer than their counterparts in clusters, then these
stars would not contribute much to the field UV flux.  This scenario
would require a ``fine-tuning'' of both secular drift velocities
and blow-out timescales for field O~stars.  Therefore we believe
that it is unlikely that deeply embedded, individual O~stars
are forming in the field regions of starbursts.

(2) If star formation occurs in both field and cluster
environments, the field IMF must differ from that found in clusters.
If the field IMF has a normal Salpeter slope, our data are consistent
with an upper mass cutoff of 30--50~$M_{\odot}$.  Alternatively,
the field IMF slope, $\alpha$, is steeper than Salpeter, with 
best fit values of $-3.0$ to $-3.5$.

(3) If star formation is occuring primarily in clusters
and associations, then the field could be composed of young,
coeval but lower mass clusters.  Our photometry and spectroscopy
both imply limits of several $100~M_{\odot}$ for groups
or clusters of young stars which could be hiding in our field spectra.
This limit is consistent with a lack of O~stars only {\it if\/}
there is an upper stellar mass limit which scales with the
total cluster mass.  So effectively, this is similar to scenario~(2).

(4) If star formation occurs primarily in clusters, but these clusters
dissolve to create the observed field, our analysis suggests that
clusters must dissolve on very rapid timescales, of order 7--10~Myr.
This is consistent with a recently presented scenario where most star
clusters, almost independent of total mass, freely expand and rapidly
disrupt.

\acknowledgements We thank the anonymous referee, whose suggestions
improved the presentation of this paper.  We are grateful for support
from NASA through grant GO-09036.01-A from the Space Telescope Science
Institute, which is operated by the AURA, Inc., for NASA under
contract NAS5-26555.  C.~A.~T.\ acknowledges support from NASA grant
NAG~58426 and NSF grant AST-0307386, and  G.~R.~M.\ acknowledges support
from NASA grant NAG5-13083.

\clearpage


\begin{deluxetable}{lccccccccc}
\tablewidth{0pc}
\tablecaption{Sample Galaxy Properties\label{table:sample}}
\tablehead{
\colhead{Galaxy} &
\colhead{Adopted Distance} &
\colhead{Area Projected in Slit\tablenotemark{a}} &
\colhead{Reference} \\
\colhead{} &
\colhead{(Mpc)} &
\colhead{(pc)} &
\colhead{}
}
\startdata
He~2-10 & \phn9\hphantom{.33} & $1100\times9\phn$ & 1 \\
MKN33 & 22\hphantom{.33} & $3150\times25$ & 2  \\
NGC~1741 & 57\hphantom{.33} & $6200\times49$ & 3 \\
NGC~3125 & 11.5\phn & $1400\times11$ & 4 \\
NGC~3310 & 13.3\phn & $2300\times18$ & 5 \\
NGC~4214 & \phn2.9\phn & $350\times3$ & 1 \\
NGC~4449 & \phn3.9\phn & $550\times4$ & 6 \\
NGC~4670 & 16\hphantom{.33} & $1900\times16$ & 4 \\
NGC~5253 & \phn3.33 & $400\times2$ & 7 \\
NGC~5996 & 47\hphantom{.33}& $6200\times49$ & 4 \\
NGC~7552 & 21\hphantom{.33} & $2500\times20$ & 4 \\
TOL1924-416 & 37\hphantom{.33} & $5000\times40$ & 4 \\
\enddata
\tablerefs{
(1)~Kobulnicky, Kennicutt, \& Pizagno 1999
(2)~Davidge 1989
(3)~Guseva, Izotov, \& Thuan 2000
(4)~Heckman et~al.\ 1998
(5)~Pastoriza et~al.\ 1993
(6)~Boeker et~al.\ 2001
(7)~Izotov \& Thuan 1999
}
\tablenotetext{a}{Based on Distance given in Column~2, assuming $25\arcsec$ 
coverage of the galaxy and given the slit width of the observations  }
\normalsize
\end{deluxetable}


\begin{deluxetable}{lccccccccc}
\tablewidth{0pc}
\tablecaption{Fraction of Far-UV Light From Clusters and the Field\label{table:frac}}
\tablehead{
\colhead{Galaxy} &
\colhead{Clusters} &
\colhead{Field}  
}
\startdata
He~2-10 & 0.64 & 0.36 \\
MKN33 & 0.41 & 0.59 \\
NGC~1741 & 0.54 & 0.46 \\
NGC~3125 & 0.76 & 0.24 \\
NGC~3310 & 0.28 & 0.72 \\
NGC~4214 & 0.77 & 0.23 \\
NGC~4449 & 0.38 & 0.62 \\
NGC~4670 & 0.68 & 0.32 \\
NGC~5253 & 0.32 & 0.68 \\
NGC~5996 & 0.59 & 0.41 \\
NGC~7552 & 0.57 & 0.43 \\
TOL1924-416 & 0.58 & 0.42 \\

\enddata
\tablecomments{Values determined by summing up the light from spatial plots 
for all clusters and comparing with the total light in the slit.  The
spatial plots are summed over 700~pixels, covering $\sim$1250--1700~\AA. Because
the slit locations were chosen to maximize the number of UV-luminous
clusters aligned in the slit, these fractoins represent {\it lower limits\/} to
the total fraction of field light in each galaxy.}
\normalsize
\end{deluxetable}


\begin{deluxetable}{lccccccccc}
\tablewidth{0pc}
\tablecaption{Archival WFPC2 Data\label{table:phot}}
\tablehead{
\colhead{Galaxy} &
\colhead{$E_{B-V}$\tablenotemark{a}} &
\multicolumn{3}{c}{Filters Exposure Time [sec]}
\\
\colhead{}  & 
\colhead{foreground} &
\colhead{F336W (U)} & 
\colhead{F555W (V)} &
\colhead{F814W (I)}
}
\startdata
NGC~4214  & 0.022 &  $2\times900$ & $2\times600$ & $2\times600$ \\
NGC~4449  & 0.019 &  $2\times260$ & $1\times200$ & $1\times200$ \\
NGC~5253  & 0.056 &  $2\times260$ & $1\times200$ & $2\times200$ \\

\enddata
\tablenotetext{a}{Foreground, Milky Way extinction from the Schlegel et~al.\ (1998) maps  }
\normalsize
\end{deluxetable}


\begin{deluxetable}{lccccccccc}
\tablewidth{0pc}
\tablecaption{Comparison of Mean Ages (Myr) of Dominant Cluster vs.\ Field Stellar Populations\label{table:ages}}
\tablehead{
\colhead{Galaxy} &
\colhead{Bright Clusters} &
\colhead{Faint Clusters}  & 
\colhead{Field}  
}
\startdata
He~2-10 & $4.5^{+0.4}_{-0.6}$  & ... & $4.6^{+0.5}_{-0.7}$  \\[3pt]
MKN33 & ... & $5.0^{+0.5}_{-0.5}$ & $6.0^{+0.5}_{-0.5}$  & \\[3pt]
NGC~1741 & $3.0^{+0.5}_{-0.6}$ & $6.0^{+0.5}_{-0.5}$ & $7.0^{+0.8}_{-0.4}$ \\[3pt]
NGC~3125 & ... & $3.0^{+0.5}_{-0.5}$ & $8.0^{+1.4}_{-0.4}$ \\[3pt]
NGC~3310 & $5.0^{+0.5}_{-0.5}$ & $7.0^{+0.8}_{-0.4}$ & $8.0^{+1.2}_{-0.3}$ \\[3pt]
NGC~4214 & $4.0^{+0.5}_{-0.5}$ & $7.0^{+0.9}_{-0.4}$  & $9.0^{+0.6}_{-1.0}$ \\[3pt]
NGC~4449 & ... & $5.0^{+0.5}_{-0.5}$ & $7.0^{+1.2}_{-0.6}$  \\[3pt]
NGC~4670 & $7.0^{+0.7}_{-0.4}$ & $5.0^{+0.6}_{-0.5}$ & $6.0^{+0.5}_{-0.5}$ \\[3pt]
NGC~5253 & ... & $4.0^{+0.4}_{-0.4}$ & $7.0^{+0.8}_{-0.4}$ \\[3pt]
NGC~5996 & $5.0^{+0.4}_{-0.4}$ & $6.0^{+0.8}_{-0.4}$ & $6.0^{+2.2}_{-0.6}$ \\[3pt]
NGC~7552 & $5.0^{+0.5}_{-0.5}$  & ... & $6.0^{+0.6}_{-0.4}$  \\[3pt]
TOL1924-416 & $1.0^{+0.5}_{-0.8}$ & ... & $7.0^{+1.0}_{-0.3}$  \\
\enddata
\tablecomments{The ages in Myr and associated uncertainties have been derived
by comparing the observations with instantaneous STARBURST99 models,
assuming a Salpeter IMF and upper and lower mass cutoffs of $100~M_{\odot}$
and $1~M_{\odot}$ respectively.  The uncertainties were determined by using
the bootstrap technique as described in Tremonti et~al.\ (2001)}
\normalsize
\end{deluxetable}



\begin{deluxetable}{lccccccccc}
\tablewidth{0pc}
\tablecaption{Comparison of Continuum Slopes in Cluster vs. Field Stellar Populations\label{table:beta}}
\tablehead{
\colhead{Galaxy} &
\multicolumn{3}{c}{FUV slope ($\beta$)\tablenotemark{a}} 
\\
\colhead{}  & 
\colhead{Bright Clusters} &
\colhead{Faint Clusters}  & 
\colhead{Field}  
}
\startdata
He~2-10 & $-1.66$  & ... & $-0.59$\\
MKN33 & ... & $-1.89$ & $-1.37$  &\\
NGC~1741 & $-1.18$ & $-1.75$ & $-2.27$\\
NGC~3125 & ... & \phs$0.12$ & $-0.84$\\
NGC~3310 & $-0.91$ & $-1.21$ & $-1.37$\\
NGC~4214 & $-1.18$ & $-2.05$  & $-0.94$\\
NGC~4449 & ... & $-3.02$ & $-2.12$\\
NGC~4670 & $-1.95$ & $-1.65$ & $-2.14$\\
NGC~5253 & ... & $-1.03$ & $-1.48$\\
NGC~5996 & $-2.02$ & $-2.42$ & $-2.74$\\
NGC~7552 & $-0.87$  & ... & $+0.11$\\
TOL1924-416 & $-2.27$ & ... & $-2.02$\\
\enddata
\tablenotetext{a}{The far UV slope, $\beta$, measured after correcting
for foreground extinction.  Typical uncertainties on $\beta$ measurements
are $\pm0.1$. }
\normalsize
\end{deluxetable}


\begin{deluxetable}{lccccccccc}
\tablewidth{0pc}
\tablecaption{Mean Star Formation Rates Derived for STIS Field Regions\label{table:sfr}}
\tablehead{
\colhead{Galaxy} &
\colhead{pixels\tablenotemark{a}} &
\colhead{$E_{B-V}$\tablenotemark{b}} &
\colhead{$L_{1500}$\tablenotemark{c}} &
\colhead{SFR\tablenotemark{d}} &
\colhead{\#O stars\tablenotemark{e}} 
}
\startdata
MKN33 & 181 & 0.18 & 4.7E38 & 1.8 &   \phn490 \\
NGC~1741 & 221 & 0.04 & 6.7E38 & 0.3    & \phn720 \\
NGC~3125 & 365 & 0.28 & 3.8E38 & 2.7   & \phn410 \\
NGC~3310 & 623 & 0.18 & 5.6E38 & 1.7   & \phn590 \\
NGC~4214 & 533 & 0.24 & 3.7E37 & 2.9   & \phn\phn40 \\
NGC~4449 & 298 & 0.11 & 5.6E36 & 0.4   & \phn\phn10 \\
NGC~4670 & 264 & 0.06 & 1.0E38 & 0.6  & \phn110 \\
NGC~5253 & 633 & 0.18 & 2.4E37 & 2.4   & \phn\phn20 \\
NGC~5996 & \phn65 & 0.00 & 2.2E38 & 0.5  & \phn230 \\
NGC~7552 & 147 & 0.41 & 1.5E39 & 8.0  & 1600 \\
TOL1924-416 & 271 & 0.08 & 9.7E38 & 0.9  & 1000 \\

\enddata
\tablenotetext{a}{Total number of pixels summed up in the spatial direction
for each field region.}
\tablenotetext{b}{Reddening derived using the starburst obscuration
curve from the difference in the model slope ($-2.5$), and the observed
slope of the field after correction for foreground reddening.  }
\tablenotetext{c}{Extinction corrected luminosity at 1500~\AA\ in
ergs~$\mbox{s}^{-1}~\mbox{\AA}^{-1}$ measured from the STIS spectra.
We use the $E_{B-V}$ values quoted in column~3.}
\tablenotetext{d}{Star formation rate in $M_{\odot}~\mbox{yr}^{-1}~\mbox{kpc}^{-2}$.  We have assumed a standard Salpeter IMF ($\alpha=-2.35$), and an
upper mass cutoff for the IMF of $100~M_{\odot}$.  The age for the continuous star formation model has been fixed at 50~Myr.  }
\tablenotetext{e}{Number of O stars predicted by STARBURST99 continuous star 
formation models (assuming a standard Salpeter IMF and 
$M_{up}=100~M_{\odot}$) to reside in the field region
given the mean SFR given in column~5.}
\normalsize
\end{deluxetable}


\begin{deluxetable}{lccccccccc}
\tablewidth{0pc}
\tablecaption{Best Fitting Field Models\label{table:field}}
\tablehead{
\colhead{Galaxy} &
\colhead{$M_{up}$\tablenotemark{a}}  & 
\colhead{$\alpha$\tablenotemark{b}}  
}
\startdata
He~2-10 &  100 &$-2.35$\\
Mkn33  & 100 &$-2.35$\\
NGC~1741 & \phn50 &$-3.0$\hphantom{5} to $-3.5$\\
NGC~3125 & \phn40 &$-3.0$\phn\\
NGC~3310 & \phn40 &$-3.0$\hphantom{5} to $-3.5$\\
NGC~4214 & \phn30 &$-3.5$\phn\\
NGC~4449 & \phn50 &$-3.0$\phn\\
NGC~4670 & 100 &$-2.35$ to $-3.0$\\
NGC~5253 & \phn50 &$-3.0$\phn\\
NGC~5996 & 100 &$-3.0$\hphantom{5} to $-3.5$\\
NGC~7552 & 100 &$-2.35$ to $-3.0$\\
TOL1924-416 & \phn50 &$-3.0$\hphantom{5} to $-3.5$\\
\normalsize
\enddata
\tablenotetext{a}{Best fit when field spectra are compared
with continuous formation STARBURST99 models which have Salpeter IMF,
$M_{low}=1~M_{\odot}$, and a variable upper mass cutoff.}
\tablenotetext{b}{Best fit to the IMF slope, $\alpha$,
in continuous formation STARBURST99 model.} 
\end{deluxetable}

\clearpage

\begin{figure}
\plotone{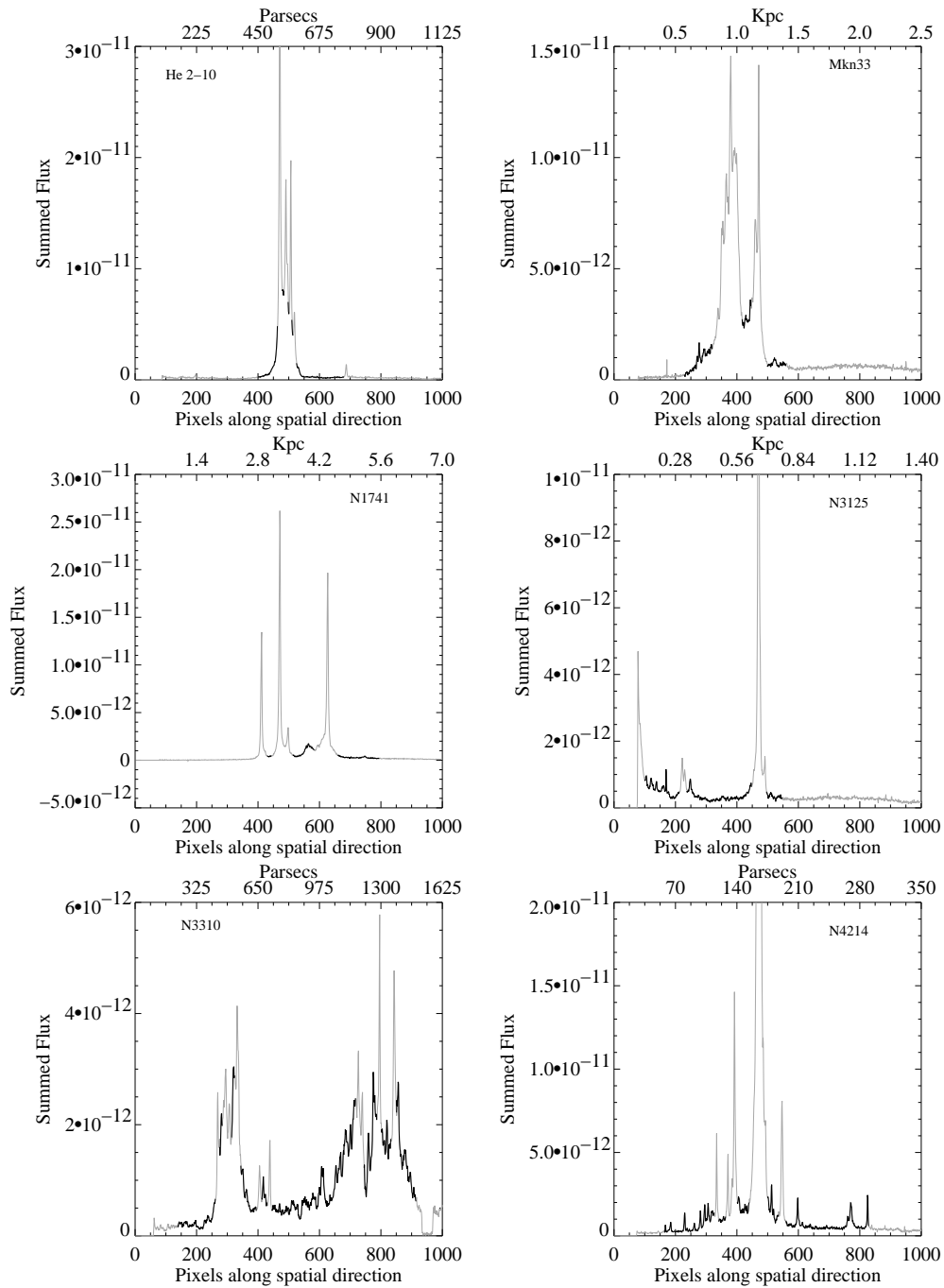}
\caption{The integrated flux (from 700 pixels, $\sim$1250--1700~\AA) as a function
of position along the slit for each target galaxy in this study.
The portions of the spectra which were used to create the composite field
spectra are plotted in black.  The STIS MAMA
detectors have a plate scale of $0.024\arcsec~\mbox{pix}^{-1}$.
\label{fig1}}
\end{figure}

\addtocounter{figure}{-1}
\begin{figure}
\plotone{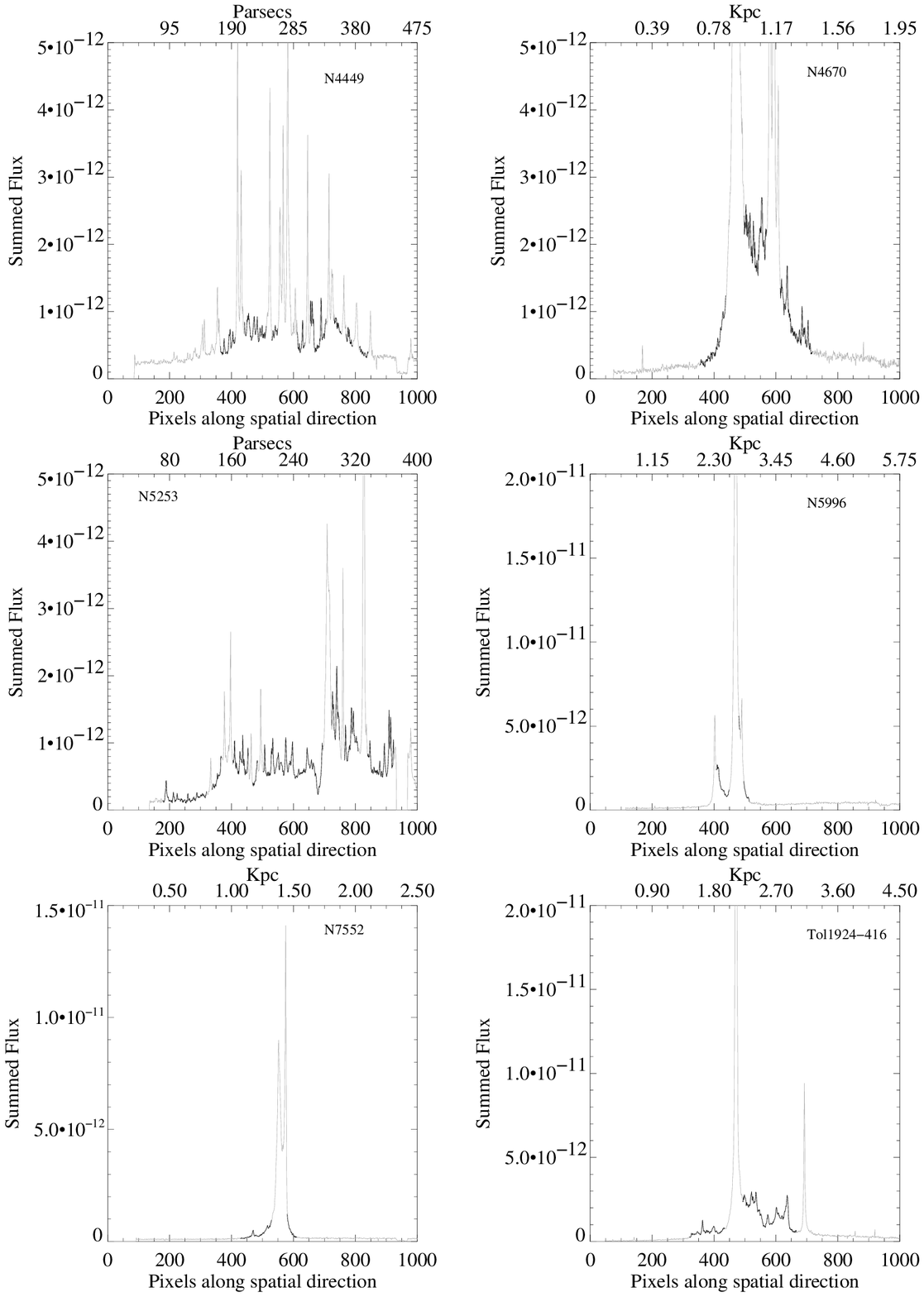}
\caption{\emph{Continued}}
\end{figure}

\begin{figure}
\epsscale{.4}
\plotone{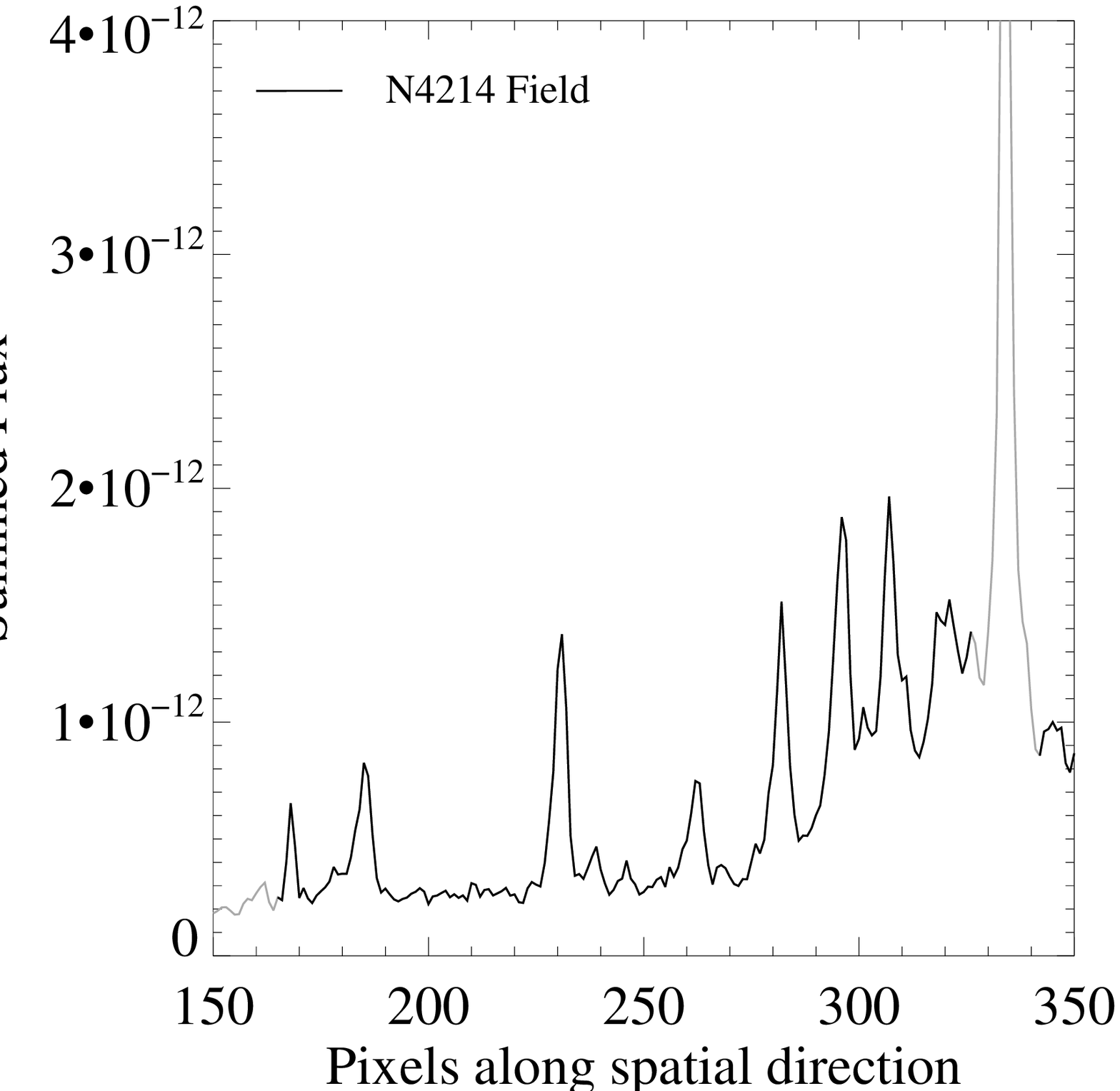}
\caption{An enlargement of the flux along the spatial direction
for a portion of NGC~4214 (in grey) containing some field regions
(in black).  This figure shows how the field light is not smooth, but
contains a number of faint peaks and valleys.
\label{fig3}}
\end{figure}

\begin{figure}
\epsscale{0.6}
\plotone{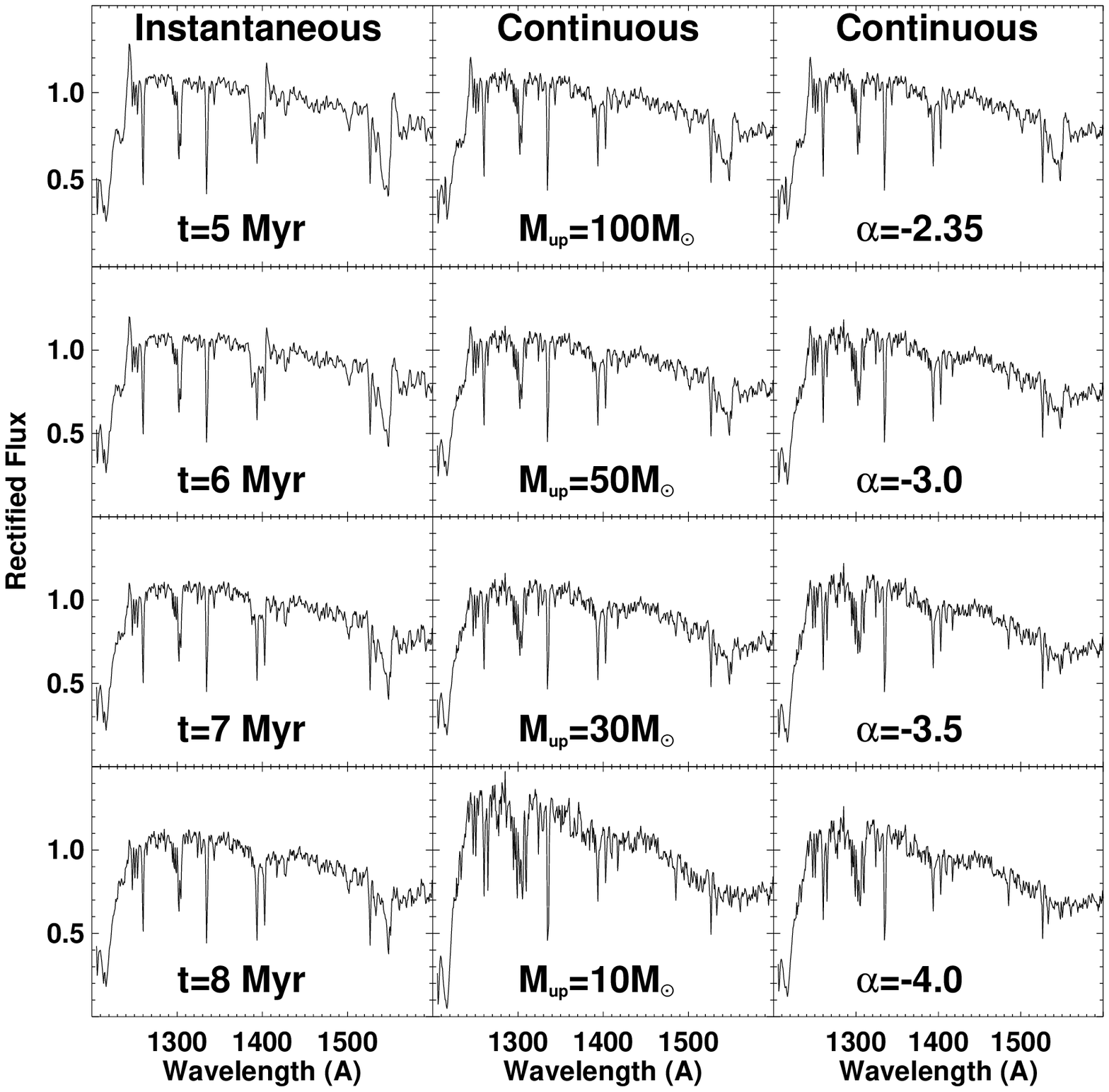}
\caption{The figure presents example spectra from various STARBURST99
models considered in this work.  The strength of the
Si~IV~$\lambda1400$ and C~IV~$\lambda1550$ wind features, plus other
diagnostics provide excellent constraints on the massive star content
in nearby galaxies.  The first column shows the effect of aging on the
UV spectrum of an instantaneous burst stellar population, with
$M_{up}=100~M_{\odot}$.  The second set of panels shows the
variations in a continuous star formation episode when the 
upper mass cutoff for the IMF is lowered from
$100~M_{\odot}$ to $10~M_{\odot}$.  And the third column shows UV
spectra for continous star formation models with different IMF slopes,
$\alpha$.  The duration for the continuous star formation shown in
columns~2 and 3 is 50~Myr.
\label{fig4}}
\end{figure}

\begin{figure}
\epsscale{1}
\plotone{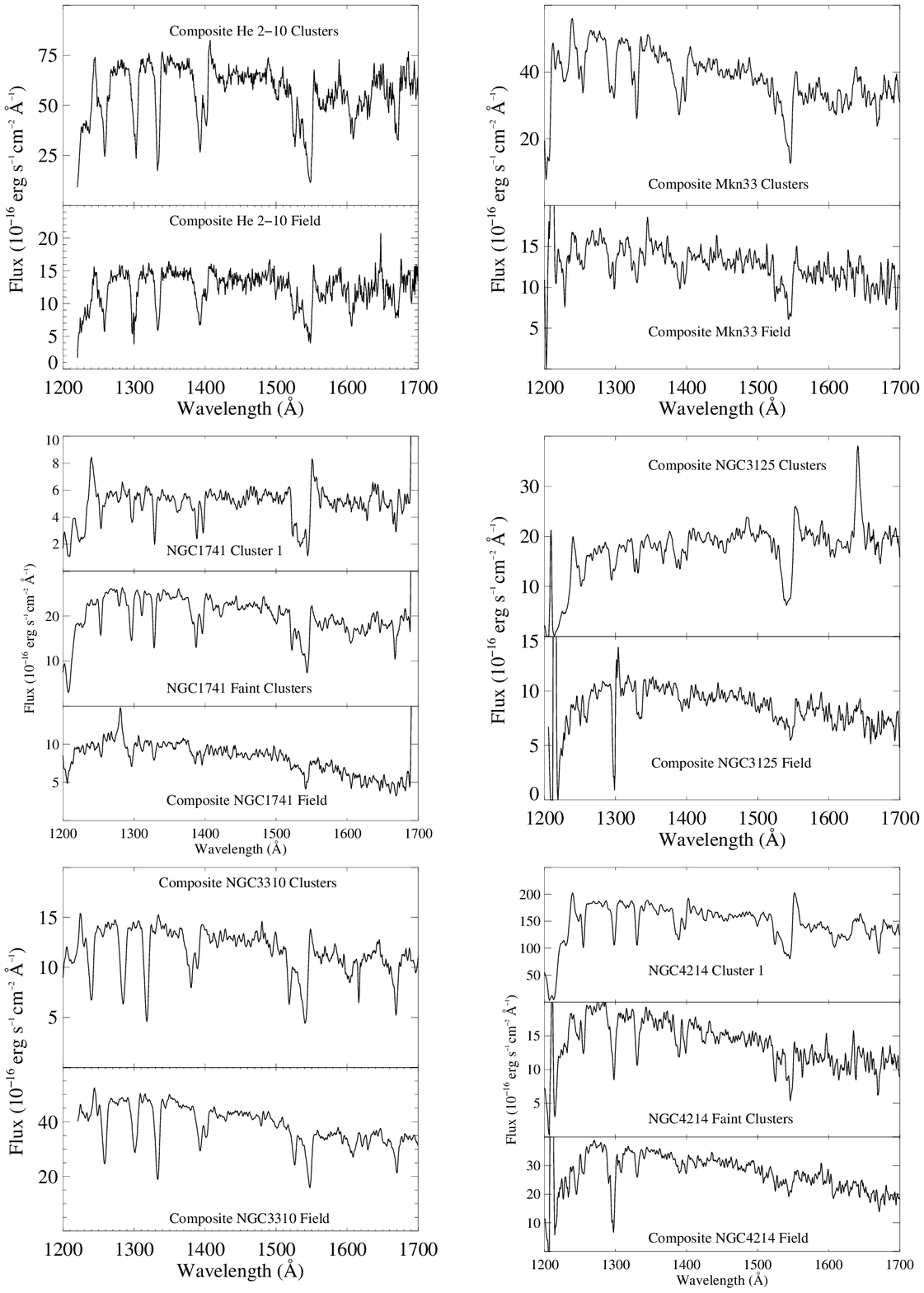}
\caption{Unweighted sum of star cluster spectra and 
summed field spectra extracted for our target galaxies.
\label{fig5}}
\end{figure}

\addtocounter{figure}{-1}
\begin{figure}
\plotone{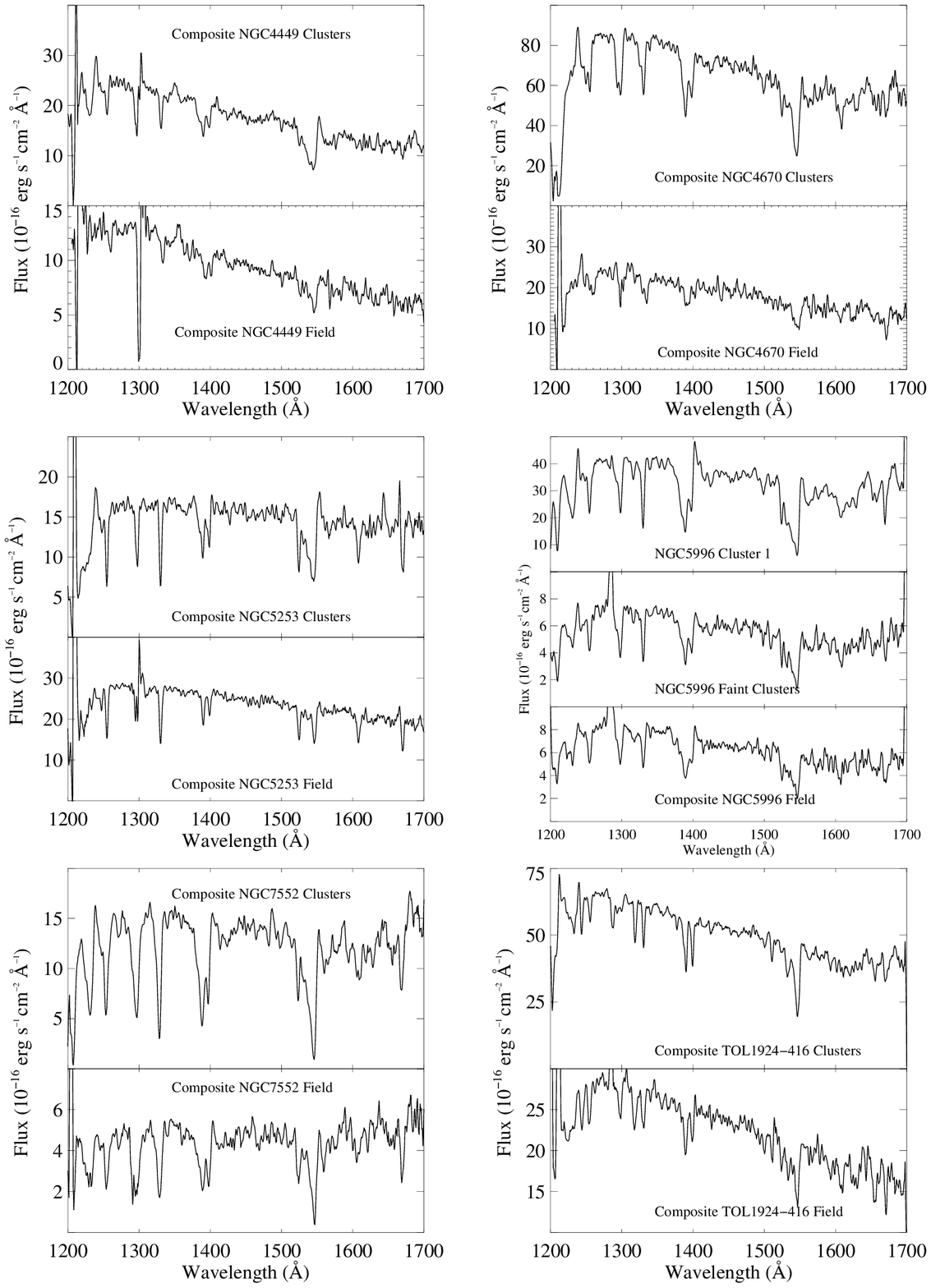}
\caption{\emph{Continued}}
\end{figure}

\begin{figure}
\epsscale{0.8}
\plotone{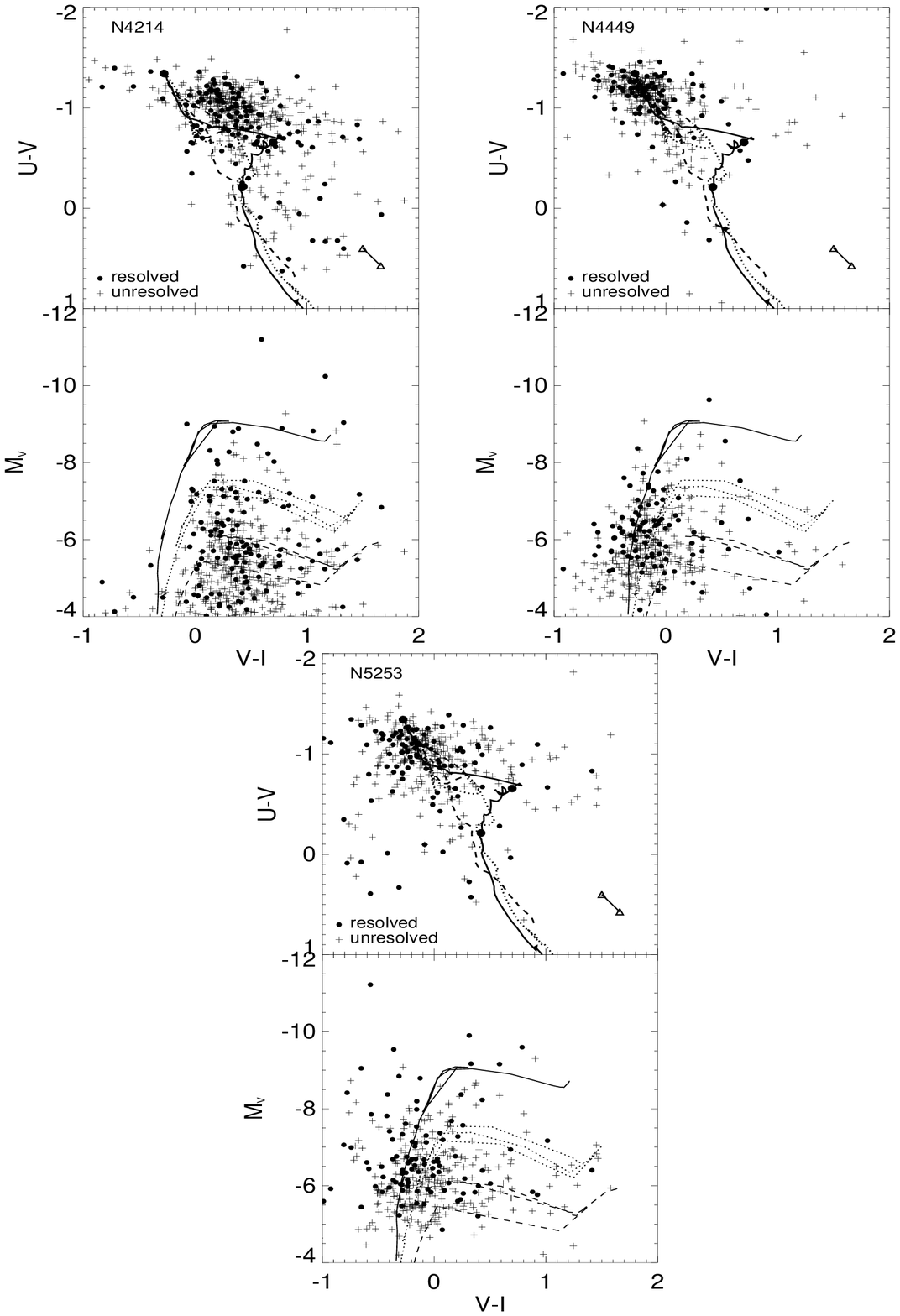}
\caption{Two$-$color and color$-$magnitude diagrams for NGC~4214,
NGC~4449, and NGC~5253 are shown, based on archival WFPC2 imaging.
The top panels show the $U-V$ vs.\ $V-I$ color-color diagrams,
corrected for foreground extinction only.  Resolved (red circles) and
unresolved (blue crosses) sources are plotted separately.  The
evolution of simple stellar populations from the Bruzual \& Charlot
(2003) models are shown for three different metallicities: solar
(solid line), $\frac{1}{5}$~solar (dotted), and $\frac{1}{50}$~solar
(dashed).  The direction of the reddening vector is shown by the
arrows, which represents $E_{B-V}=0.1$.  Ages of $10^6$, $10^7$, and
$10^8$ are shown as black filled circles along the cluster models,
starting from the upper left.  In the bottom panels, we show the $V$
vs.\ $V-I$ color magnitude diagrams.  Overplotted are $30~M_{\odot}$,
$15~M_{\odot}$, and $9~M_{\odot}$ Padova stellar evolution tracks 
(Fagotto et~al.\ 1994a,b), with $\frac{1}{5}$~solar metallicity.
\label{fig7}}
\end{figure}

\begin{figure}
\epsscale{.5}
\plotone{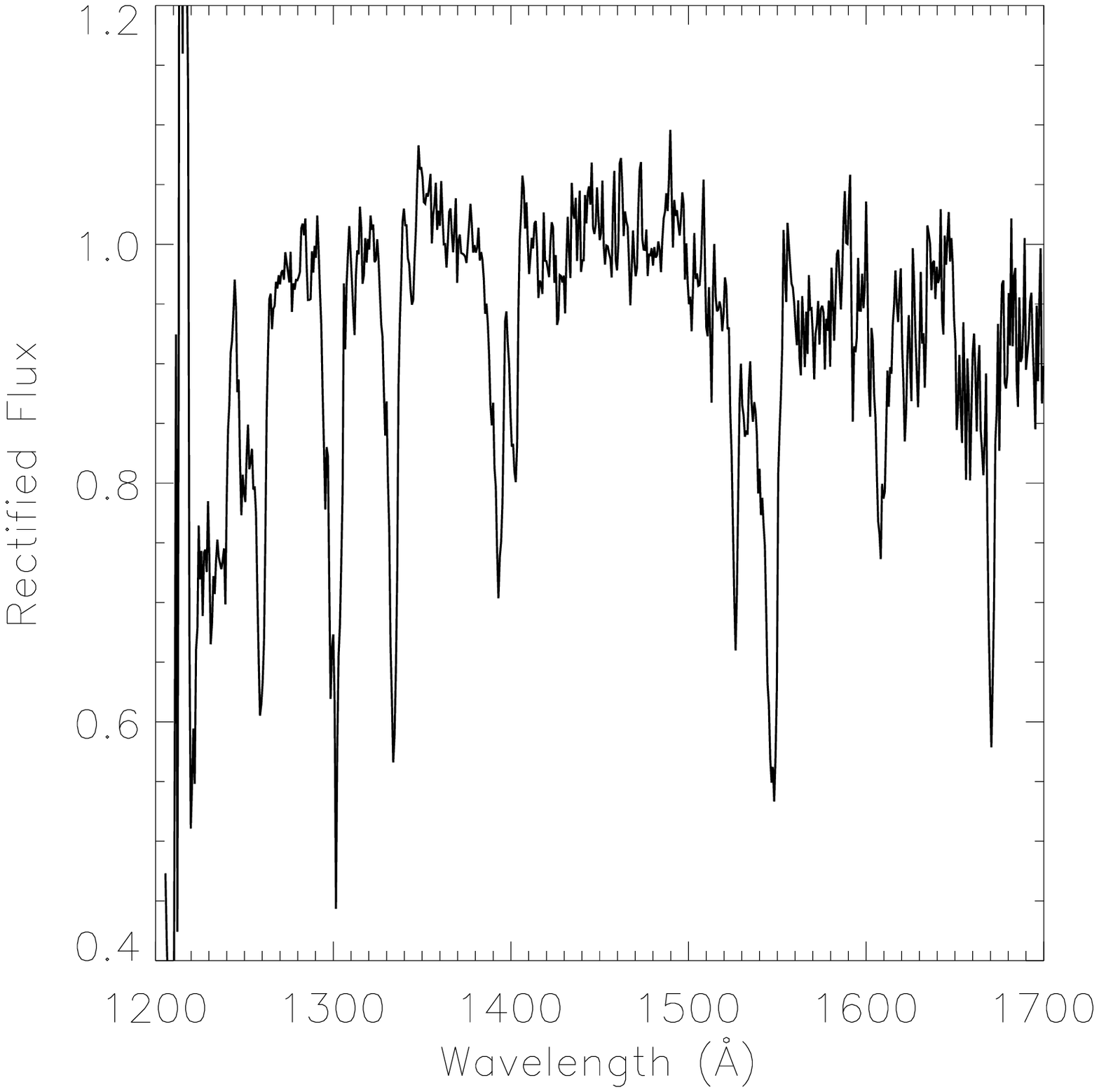}
\caption{Rectified spectrum of the unweighted sum of field regions for
six low metallicity galaxies which have B-star dominated field spectra
(MKN33, NGC~4214, NGC~4449, NGC~4670, NGC~5253, and NGC~1741).
\label{fig8}}
\end{figure}

\end{document}